\documentclass[letterpaper,10pt,journal,twoside]{IEEEtran}
\IEEEoverridecommandlockouts
\usepackage{amsmath,amssymb,amsfonts,bm}
\usepackage[colorlinks, linkcolor=magenta, anchorcolor=green, citecolor=blue]{hyperref}
\usepackage{xcolor}
\usepackage{algorithmic}
\usepackage{caption}
\usepackage{array}
\usepackage{textcomp}
\usepackage{stfloats}
\usepackage{url}
\usepackage{verbatim}
\usepackage{algorithmic}
\usepackage[linesnumbered,ruled]{algorithm2e}

\newcommand*{\circled}[1]{\lower.7ex\hbox{\tikz\draw (0pt, 0pt)%
    circle (.5em) node {\makebox[1em][c]{\small #1}};}}
\usepackage[square, numbers]{natbib}
\usepackage{xcolor} 
\newcommand{\cmark}{\textcolor[rgb]{0,0.5,0}{\ding{51}}} 
\newcommand{\xmark}{\textcolor[rgb]{0.6,0,0}{\ding{55}}} 
\usepackage{bigstrut,multirow,rotating}
\usepackage{booktabs}
\usepackage{multirow}
\usepackage{array}
\usepackage{multirow}
\usepackage{rotating}
\usepackage{graphicx,pifont}
\usepackage[normalem]{ulem}

\usepackage{makecell}

\usepackage{wasysym}
\newcommand{\circnum}[1]{\textcircled{\scriptsize #1}}

\let\oldding\ding
\renewcommand{\ding}[2][1]{\scalebox{#1}{\oldding{#2}}}
\def\BibTeX{{\rm B\kern-.05em{\sc i\kern-.025em b}\kern-.08em
    T\kern-.1667em\lower.7ex\hbox{E}\kern-.125emX}}

\begin{document}

\title{
\huge
ACE-GNN: Adaptive GNN Co-Inference with System-Aware Scheduling in Dynamic Edge Environments}

\author{Ao~Zhou,~Jianlei~Yang,~\IEEEmembership{Senior~Member,~IEEE},~Tong~Qiao,~Yingjie Qi,~Xinming Wei,~Cenlin Duan,\\Weisheng~Zhao,~\IEEEmembership{Fellow,~IEEE}, Chunming~Hu
\thanks{Manuscript received on April, revised on July 2025, and accepted on September 2025.
This work is supported in part by the National Natural Science Foundation of China (Grant No. 62072019), the Beijing Natural Science Foundation (Grant No. L243031) and the National Key R\&D Program of China (Grant No. 2023YFB4503704 and 2024YFB4505601).
\textit{Corresponding author is Jianlei Yang.}}
\thanks{A. Zhou and C. Hu are with School of Software, Beihang University, Beijing 100191, China.}
\thanks{J. Yang, T. Qiao, Y. Qi and X. Wei are with School of Computer Science and Engineering, Beihang University, Beijing 100191, China, and Qingdao Research Institute, Beihang University, Qingdao 266104, China. Email: \url{jianlei@buaa.edu.cn}.}
\thanks{C. Duan and W. Zhao is with School of Integrated Circuits and Engineering, Beihang University, Beijing 100191, China.
}

}

\maketitle
\bstctlcite{IEEEexample:BSTcontrol}

\begin{abstract}
The device-edge co-inference paradigm effectively bridges the gap between the high resource demands of Graph Neural Networks (GNNs) and limited device resources, making it a promising solution for advancing edge GNN applications.
Existing research enhances GNN co-inference by leveraging offline model splitting and pipeline parallelism (PP), which enables more efficient computation and resource utilization during inference.
However, the performance of these static deployment methods is significantly affected by environmental dynamics such as network fluctuations and multi-device access, which remain unaddressed.
We present ACE-GNN, the first \underline{A}daptive GNN \underline{C}o-inference framework tailored for dynamic \underline{E}dge environments, to boost system performance and stability.
ACE-GNN achieves performance awareness for complex multi-device access edge systems via system-level abstraction and two novel prediction methods, enabling rapid runtime scheme optimization.
Moreover, we introduce a data parallelism (DP) mechanism in the runtime optimization space, enabling adaptive scheduling between PP and DP to leverage their distinct advantages and maintain stable system performance.
Also, an efficient batch inference strategy and specialized communication middleware are implemented to further improve performance.
Extensive experiments across diverse applications and edge settings demonstrate that ACE-GNN achieves a speedup of up to $12.7\times$ and an energy savings of $82.3\%$ compared to GCoDE, as well as $11.7\times$ better energy efficiency than Fograph.

\end{abstract}


\section{Introduction}\label{sec:introduction}

\begin{figure}[t]
    \centering
    \includegraphics[width = 1\linewidth]{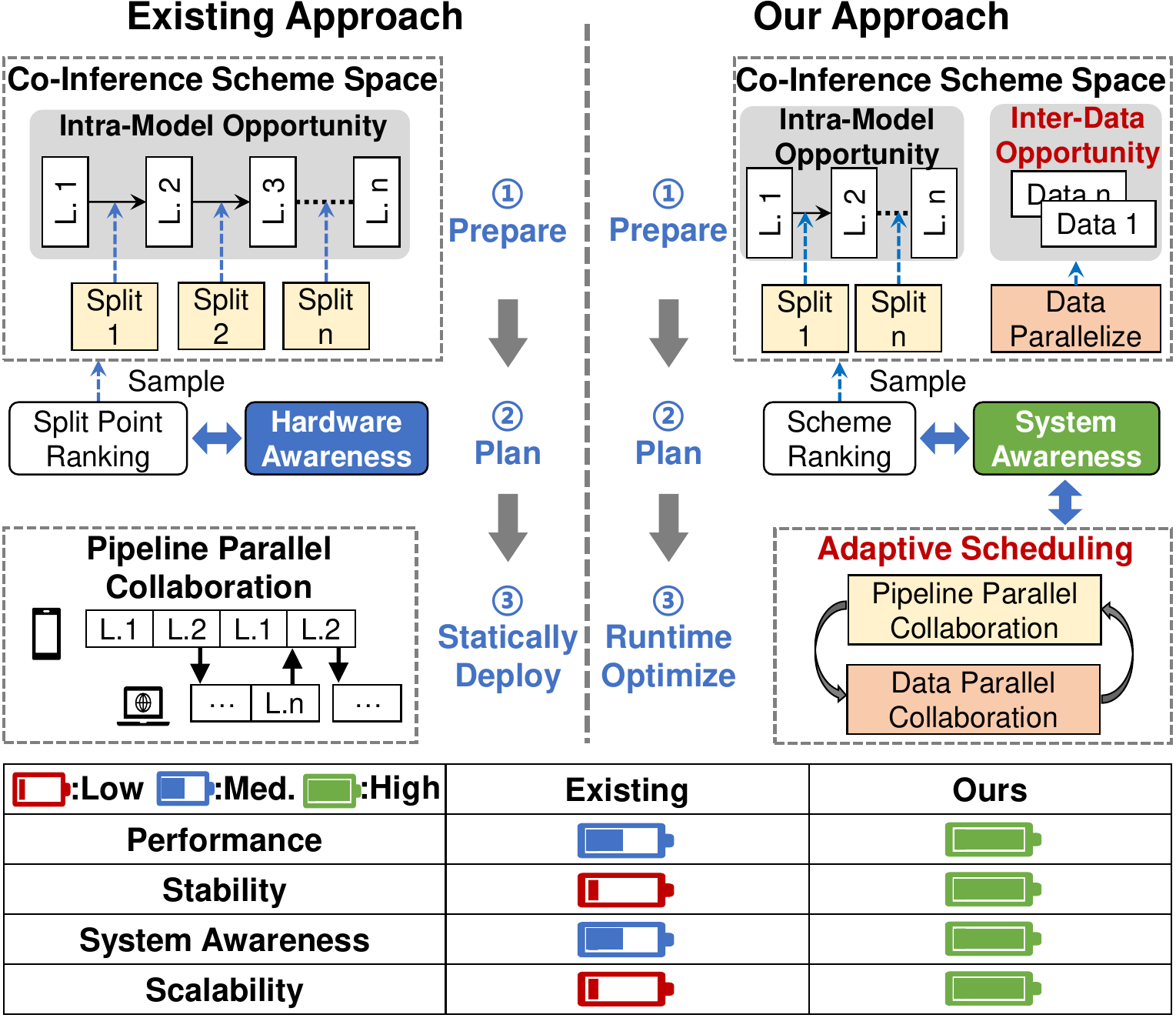}
    \caption{Device-Edge Co-Inference for GNNs.}
    \label{fig:approach_compare}
    \vspace{-9pt}
\end{figure}
\IEEEPARstart{W}ith deep learning breakthroughs continuing, the demand for deploying AI applications on edge devices for real-time data processing has rapidly increased~\cite{sheng2022larger,xue2023sugar}.
However, limited resources on edge devices pose significant challenges in meeting latency requirements when deploying complex models~\cite{zhou2024hgnas}. 
For example, Graph Neural Networks (GNNs) have achieved state-of-the-art (SOTA) performance in handling irregular data, becoming a popular strategy for applications such as point cloud processing~\cite{tailor2021towards}, natural language processing~\cite{chen2023multivariate}, and Internet of Things (IoT)~\cite{zeng2023serving}.
Despite their impressive achievements, GNNs often suffer from prohibitive inference costs, limiting their potential in real-world edge applications~\cite{zhang2021g, you2022gcod}.
This limitation is demonstrated by deploying GNNs tailored by HGNAS~\cite{zhou2023hardware} on a Raspberry Pi 4B, achieving only $4.1$ fps in point cloud processing, which is far below real-time requirements.

To tackle the inefficiency of GNNs in edge scenarios, researchers are increasingly focusing on the emerging device-edge co-inference paradigm~\cite{shao2021branchy, zhou2024graph}.
Unlike traditional model co-inference, GNNs pose unique challenges in co-inference design, yet research in this area remains scarce.
First, the GNN inference process exhibits the phenomenon of data amplification, leading to high communication overhead.
Meanwhile, GNN operations are hardware-sensitive, making simple model splitting insufficient to accommodate device-edge heterogeneity.
To reduce communication overhead, Branchy~\cite{shao2021branchy} introduces a model partitioning strategy that splits an existing GNN model into device-execute and edge-execute parts, complemented by a data compression scheme to further reduce transmission costs.
Expanding this concept, GCoDE~\cite{zhou2024graph} employs hardware-aware architecture search to design GNNs with embedded split points for addressing system heterogeneity.
Additionally, model splitting enables intra-model parallelism, improving performance through pipeline processing across hierarchies.

While the performance of current GNN co-inference approaches is impressive, it is insufficient to meet the efficiency and stability demands in real-world edge applications.
This limitation arises from the reliance on static scheme designs, which fail to address a key characteristic of real edge environments: \textbf{dynamism}.
Such dynamics can severely impact co-inference system performance~\cite{zhuang2024nebula}.
In this paper, we focus on three key factors: (1) fluctuations in network speed, (2) variations in edge server workloads with multi-device access, and (3) changes in available system resources.
In practice, we find that as network degradation and edge workload increase, the optimal preset co-inference scheme often fails to guarantee real-time task processing.
Moreover, addressing the underutilization of idle devices is crucial for improving system performance.
Additionally, the benefits of pipeline parallelism depend on model and data characteristics, and may become ineffective when it is hard to identify a split point that balances computation and communication.
Besides, system performance awareness for GNN co-inference is more challenging than traditional models, making it a major issue in adaptive system design.
It is affected by model structure, graph topology, and system heterogeneity, and is further complicated by multi-device access.

To address the above challenges, we propose ACE-GNN, the first GNN co-inference framework tailored for dynamic edge environments. Unlike prior work such as GCoDE~\cite{zhou2024graph}, which focuses on static model-level architecture–partition co-design, ACE-GNN supports system-level adaptation through real-time scheduling, system awareness, and collaborative mechanisms across multiple edge devices.
As shown in Fig.~\ref{fig:approach_compare}, ACE-GNN integrates data and pipeline parallelism with system-aware adaptive scheduling, outperforming existing methods in performance, stability, and scalability.
ACE-GNN’s contribution goes beyond simply leveraging device-edge hierarchies for GNN inference acceleration, and instead addresses these challenges at four levels.
First, from a collaborative perspective, data parallelism is introduced as a candidate scheme to mitigate the drawbacks of GNN model splitting in dynamic edge environments.
Second, a novel and effective system-awareness mechanism tailored for multi-device edge environments is designed to enable accurate runtime evaluation of co-inference schemes.
This complex system awareness is built on the concept that \textbf{the awareness approach aims to identify the more efficient scheme at runtime rather than providing precise values}.
Such an idea is well applied through system-level abstraction and relative performance prediction, effectively simplifying the awareness problem.
Third, a hierarchical optimization algorithm is applied to rapidly identify the optimal scheme at runtime, and an efficient batch inference strategy is designed to optimize edge resource utilization.
Finally, an efficient co-inference engine and communication middleware are integrated to support adaptive scheduling.
Moreover, the overhead of adaptive optimization is negligible, as the primary system awareness runs on idle edge threads, with each evaluation taking less than $2$ ms.
Extensive evaluations demonstrate that ACE-GNN outperforms all baselines.
The main contributions are summarized as follows:
\begin{itemize}
    \item We propose ACE-GNN, the first adaptive GNN co-inference framework for dynamic edge environments, supporting rapid runtime optimization to maintain stable performance.
    \item We propose a novel system-aware method for efficient runtime scheme evaluation, which shows up to $97.3\%$ accuracy in relative performance prediction. To the best of our knowledge, ACE-GNN is also the first to achieve performance awareness in GNN co-inference systems involving multiple edge devices.
    \item We explore the drawbacks of pipeline parallelism in GNN co-inference and present a data parallelism strategy to resolve them, ensuring a balance between communication and computation.
    \item We implement ACE-GNN and evaluate it across various GNN models, applications, and system configurations. Experimental results show up to $12.7\times$ speedup and $82.3\%$ on-device energy savings over SOTA GNN co-inference frameworks.
\end{itemize}
\section{Preliminaries and Motivations}\label{sec:motivation}
\subsection{GNN Co-Inference}

\begin{figure}[t]
    \centering
    \includegraphics[width = 1\linewidth]{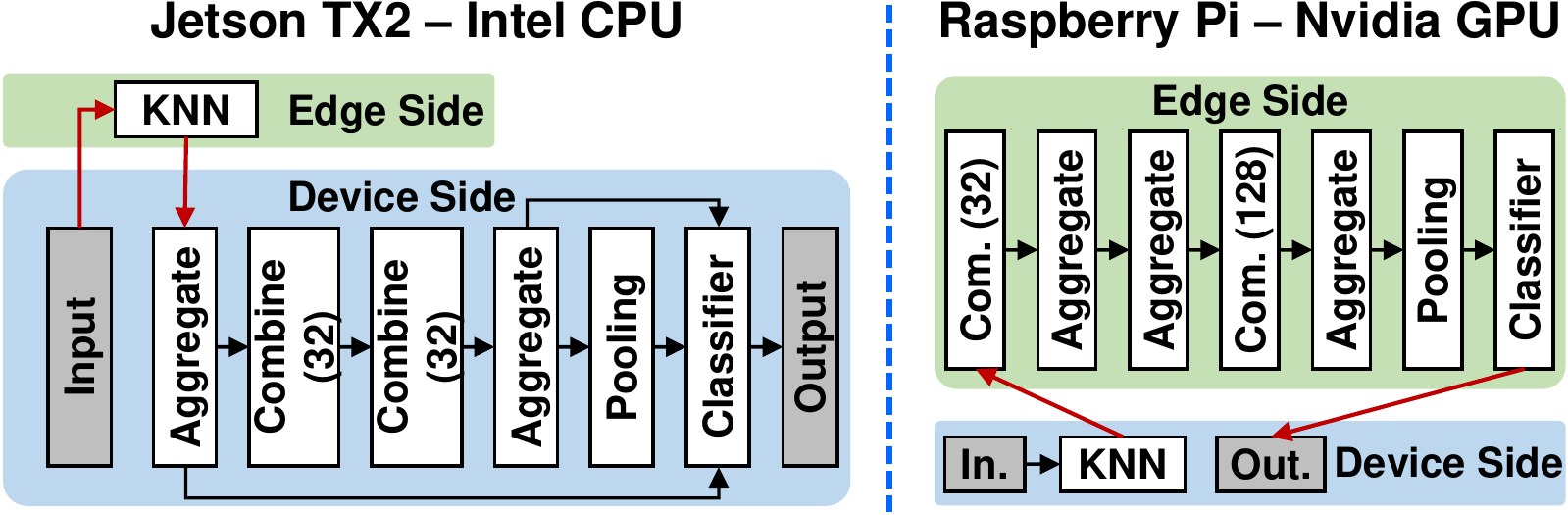}
    \caption{Existing GNN co-inference designs from GCoDE, leveraging system heterogeneity.}
    \label{fig:gcode}
    \vspace{-6pt}
\end{figure}
Device-edge co-inference is an emerging deployment paradigm for edge AI applications that alleviates resource constraints on edge devices.
In this paradigm, model partitioning is widely adopted to distribute inference tasks across edge devices and edge servers, while intermediate feature data is transmitted via wireless networks.

Branchy~\cite{shao2021branchy} was the first to explore GNN device-edge co-inference and identified the data amplification phenomenon.
It splits the GNN model to reduce communication overhead, but neglects hardware diversity, leading to limited performance gains.
Taking point cloud processing as an example, GNN inference typically involves three main operations: \textit{Sample}, \textit{Aggregate}, and \textit{Combine}, each with distinct hardware sensitivities~\cite{zhou2024hgnas}.
For instance, the memory-intensive KNN sampling operation is a primary bottleneck on GPUs, but not on CPUs~\cite{zhou2023hardware}.
To exploit hardware heterogeneity, GCoDE~\cite{zhou2024graph} introduces a hardware-aware framework for automated GNN architecture–partition co-design. It employs a model-level performance predictor to jointly search for architectures and partition points under fixed system settings. 
As shown in Fig.~\ref{fig:gcode}, GCoDE statically allocates operations (e.g., assigning memory-intensive KNN sampling to the CPU) based on hardware characteristics, enabling heterogeneous acceleration. 
To enable large-scale graph inference, Fograph~\cite{zeng2022fograph} adopts subgraph partitioning to support GNN co-inference across multiple edge devices.

While these works consider GNN computation and communication characteristics, they remain static and lack runtime adaptability.
Table~\ref{tab:gcode-compare} summarizes the supported features of existing GNN co-inference frameworks. 
GCoDE relies on pipeline parallelism with a single edge device, and its system awareness is limited to bandwidth-based partition switching.
Its partitioning scheme is predetermined, does not support concurrent multi-device collaboration, and is not scalable under dynamic system conditions such as device contention or network fluctuation.
Fograph, though scalable, provides no runtime scheduling or system-awareness mechanisms.
In contrast, our framework targets system-level runtime adaptation in dynamic edge environments. 
It supports concurrent collaboration across multiple devices and introduces new collaboration strategies and a lightweight system-awareness method for efficient runtime scheme evaluation.

\begin{table}[t]
\centering
\caption{Supported feature comparison of related works}
\label{tab:gcode-compare}
\renewcommand{\arraystretch}{1.3}
\resizebox{1\linewidth}{!}{%
\begin{tabular}{lcccc}
\toprule
\textbf{Features} & \textbf{Ours} & \makecell{\textbf{GCoDE}\\\cite{zhou2024graph}} & \makecell{\textbf{Branchy}\\\cite{shao2021branchy}} & \makecell{\textbf{Fograph}\\\cite{zeng2022fograph}} \\
\midrule
Multi-Parallelism & \cmark & \xmark & \xmark & \xmark \\
Multi-Device Scalability & \cmark & \xmark & \xmark & \cmark \\
Runtime Adaptive Scheduling & \cmark & $\circ$ & \xmark & \xmark \\
System Awareness & \cmark & $\circ$ & \xmark & \xmark \\
Architecture Exploration & \xmark & \cmark & \xmark & \xmark \\
Edge Deployment & \cmark & \cmark & \cmark & \cmark \\
\bottomrule
\end{tabular}
}
\vspace{0.5em}
\footnotesize{
\cmark: supported; \xmark: not supported; $\circ$: partially supported.
}
\vspace{-9pt}
\end{table}

\subsection{Observations and Motivations}

In this section, we present three key motivations for adaptive system design in GNN co-inference, drawing on prior research and real-world deployment observations.
These motivations stem from typical edge computing conditions, including network variability~\cite{odema2021lens}, concurrent device access~\cite{jin2024task}, and resource underutilization~\cite{ye2024asteroid}, which are central challenges addressed in this work.

\begin{figure}[t]
    \centering
    \includegraphics[width = 1\linewidth]{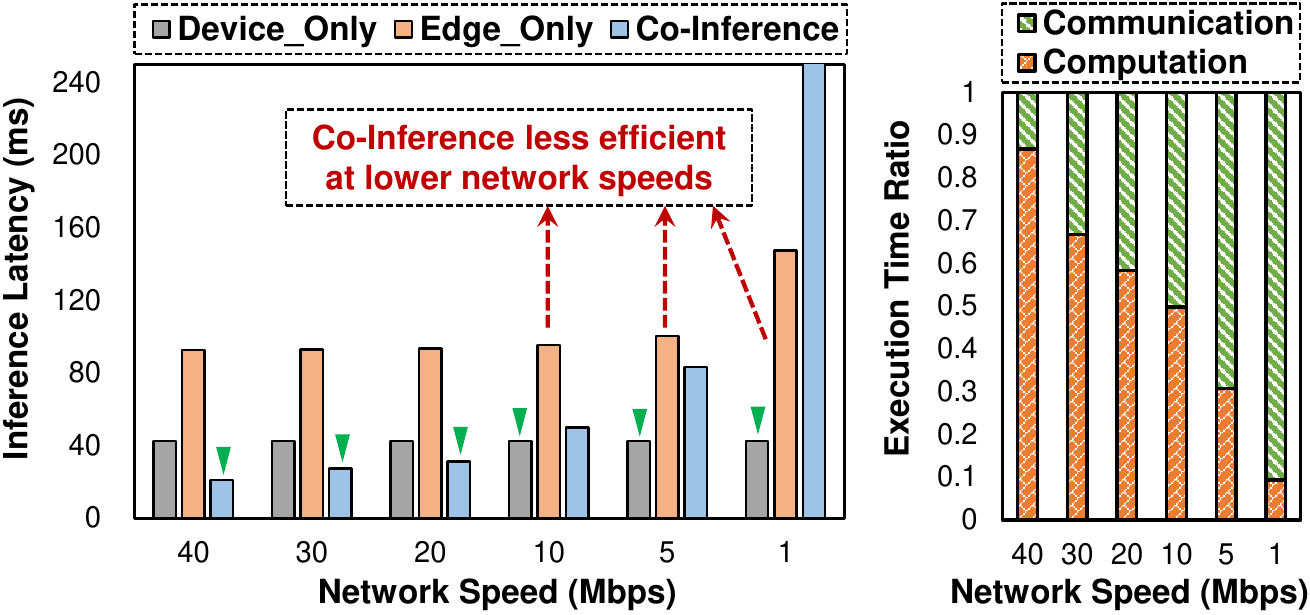}
    \caption{Comparison of GNN inference performance under varying network conditions on ModelNet40~\cite{wu20153d}, including Device-Only, Edge-Only, and Co-Inference modes on Jetson TX2 and Intel CPU.}
    \label{fig:motivation1}

\end{figure}

\textbf{Motivation 1: Diminished Pipeline Efficiency in Unstable Network Conditions.}
Existing GNN co-inference designs primarily adopt model partitioning, utilizing pipeline parallelism between edge devices and the edge server to improve inference efficiency.
However, these benefits can vary significantly.
As shown in Fig.~\ref{fig:motivation1}, the GCoDE-designed GNN co-inference scheme does not consistently achieve optimal performance under varying network conditions.
Under poor network conditions, on-device inference or offloading inference tasks to the edge server is preferable.
Examining the co-inference process reveals that pipeline parallelism provides significant performance gains at high network speeds, covering the communication overhead, but this is reversed at lower speeds.
This is because the initial node feature of the point cloud data is only three-dimensional, and model splitting causes data amplification.
Therefore, the benefits of pipeline parallelism depend on data characteristics, model structure, and network conditions, highlighting the need for new mechanisms and runtime optimizations.

\begin{figure}[t]
    \centering
    \includegraphics[width = 1\linewidth]{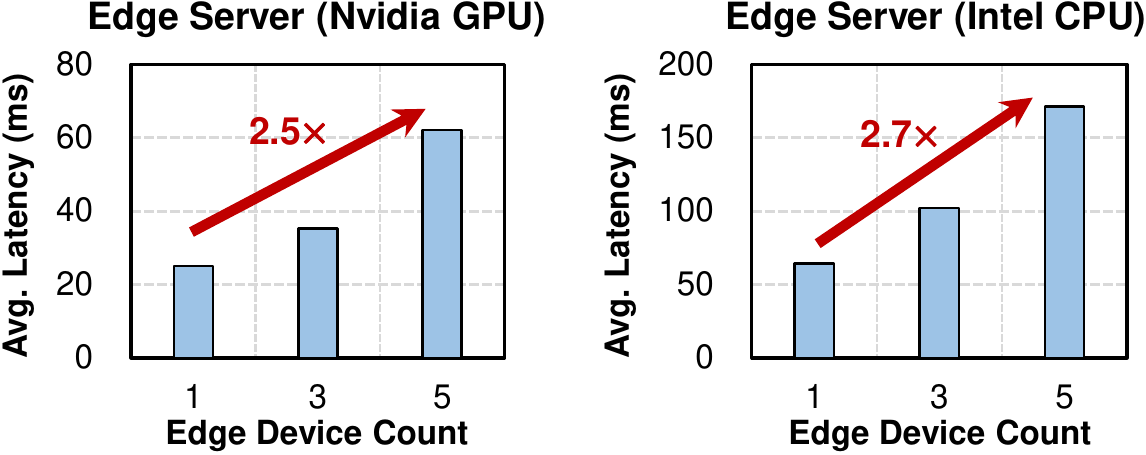}
    \caption{System efficiency under concurrent access from multiple edge devices on point cloud processing, with Raspberry Pi 4B serving as edge devices.}
    \label{fig:motivation2}
    \vspace{-9pt}

\end{figure}

\textbf{Motivation 2: Severe Edge Workload Caused by Multi-Device Concurrent Access.}
In edge environments, it is common for multiple edge devices to access the edge server for co-inference requests, presenting challenges to system performance~\cite{shi2022multiuser}.
Such concurrent co-inference imposes dual pressures of communication and computation on the edge server, resulting in reduced processing efficiency.
Fig.~\ref{fig:motivation2} shows that when five edge devices use the GCoDE co-inference scheme to connect to a single edge server, the average inference speed drops by $2.5\times$ with the Nvidia GPU and $2.7\times$ with the Intel CPU.
These issues highlight the limitations of static deployments in GNN co-inference solutions, pointing to the need for better edge server utilization and adaptive runtime scheduling.

\textbf{Motivation 3: Opportunity for Performance Gains through Leveraging Idle Resources.}
Unlike the previously discussed scenario of concurrent multi-device access for co-inference, connected edge devices may not always require it and remain idle in practice.
Leveraging these idle edge devices offers a chance to improve system performance and handle larger graphs~\cite{zeng2023serving, zeng2022gnn}.
Fograph~\cite{zeng2022fograph} utilizes multiple edge devices to construct a distributed system, enabling real-time GNN inference on graphs with thousands of nodes through graph partitioning.
Thus, runtime awareness and adaptive task scheduling mechanisms are essential for maximizing system resource utilization.

\begin{figure}[t]
    \centering
    \includegraphics[width = 1\linewidth]{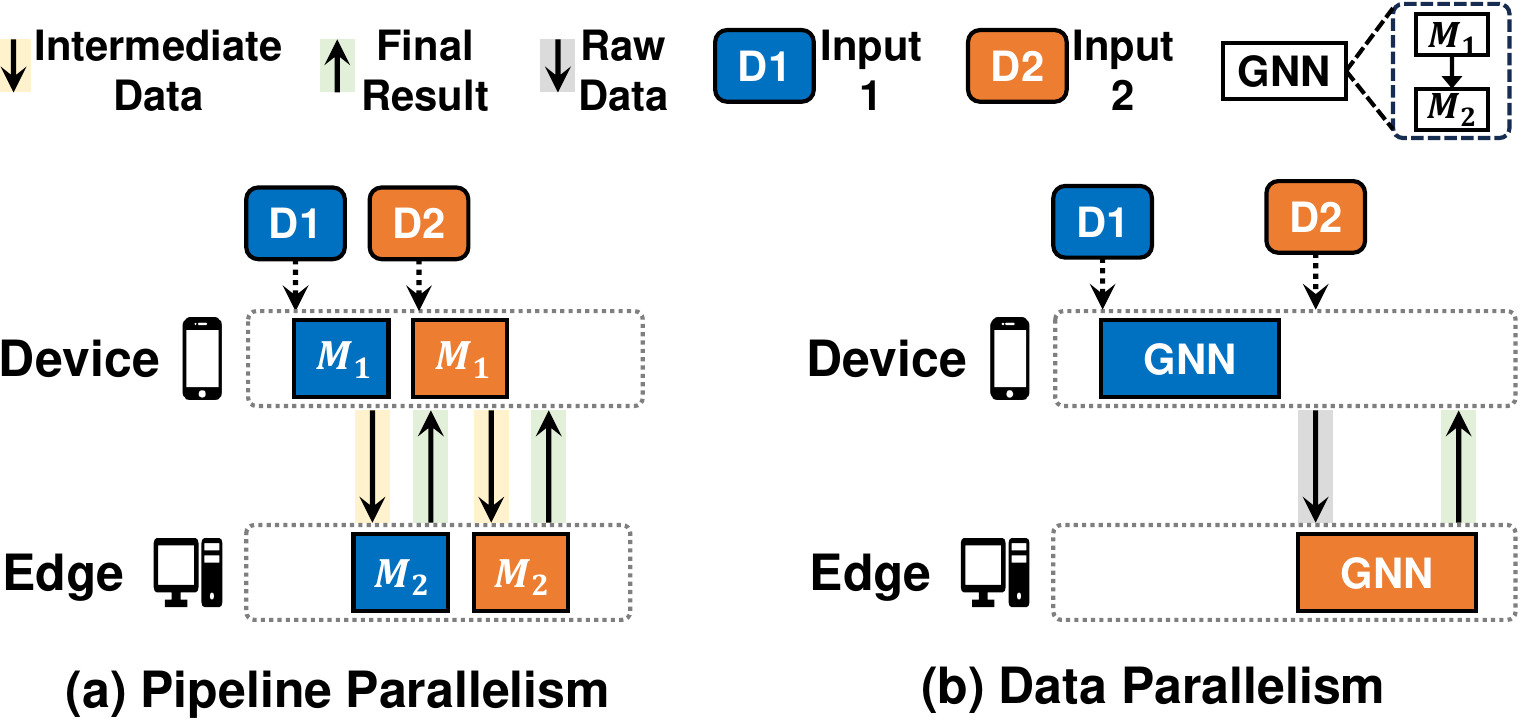}
    \caption{Illustration of pipeline parallelism (PP) and data parallelism (DP) in GNN device-edge co-inference. PP splits model stages across devices for pipelined processing, while DP executes multiple inputs in parallel using replicated models.}
    \label{fig:parallesim}

\end{figure}

\begin{table}[t]
\small
\centering
\caption{Comparison of communication volume in GNN co-inference.}
\label{tab:insight}
\renewcommand{\arraystretch}{1.3} 
\resizebox{0.85\linewidth}{!}{%
\begin{tabular}{ccccc}
\hline
\multirow{2}{*}{Volume} & \multicolumn{2}{c}{ModelNet40} & \multicolumn{2}{c}{Yelp} \\ \cline{2-5} 
                        & DGCNN          & GCoDE         & GCN         & GAT        \\ \hline
PP (KB)                 & 24.2           & 332.0         & 1154.2      & 5529.2     \\
DP (KB)                 & 12.2           & 12.2          & 4396.1      & 4396.1     \\ \hline
\end{tabular}%
}

\end{table}

\subsection{Key Insight: Adaptive Optimization of Data and Pipeline Parallelism}

The above analysis motivates us to introduce new parallel mechanisms and runtime optimization for GNN co-inference in dynamic edge environments.

Inspired by \cite{ye2024asteroid} and the specific characteristics of GNN workloads, we propose a hybrid approach that combines data parallelism (DP) and pipeline parallelism (PP) to adapt to system dynamics. As shown in Fig.~\ref{fig:parallesim}, PP partitions the model between the device and the server to process different layers of the same input concurrently, whereas DP distributes independent inputs across available computing nodes. This hybrid scheme enables new data to bypass occupied devices and be directly processed by the edge server, mitigating the intermediate data amplification commonly seen in PP.
As shown in Tab.~\ref{tab:insight}, DP significantly reduces communication overhead compared to PP for ModelNet40~\cite{wu20153d}.
For the Yelp~\cite{rayana2015collective} dataset with larger feature dimensions, GCN's~\cite{kipf2016semi} lower-dimensional hidden layers benefit from PP, while GAT's~\cite{velivckovic2018graph} multi-head attention increases the intermediate data volume of PP.
Combining this observation with the previous analysis suggests that DP may offer communication advantages, while PP can leverage heterogeneous acceleration for improved computation and reduce edge workload.
Both benefits depend on data characteristics, model structure, and runtime environment factors like network conditions and edge workload.
Therefore, a system-aware optimization method for adaptive scheduling across collaborative schemes is warranted.

\subsection{Technical Challenges}

While the combination of DP and PP in GNN co-inference offers potential to balance communication and computation, achieving runtime optimization and adaptivity in dynamic edge environments remains challenging.

\textbf{System-Awareness Level: Interference from Device Concurrency, Task Variability, and System Heterogeneity.}
Accurately estimating the runtime performance of co-inference schemes is essential for adaptive optimization. However, edge environments are inherently dynamic: concurrent device requests, input variability, and hardware heterogeneity jointly affect system behavior and hinder lightweight performance estimation. Although GCoDE~\cite{zhou2024graph} introduces a model-level latency predictor, it is hardware-specific and cannot address the complexities of multi-device systems. Achieving low-cost yet accurate system awareness under such interference remains a key challenge.

\textbf{Runtime-Scheduling Level: Large Strategy Space and Inefficient Multi-Device Scheduling.}
Adapting to dynamic edge environments requires selecting co-inference strategies based on real-time workloads and device availability. However, as the number of GNN layers and participating edge devices increases, the scheduling space grows exponentially. For instance, an $n$-layer model and $m$ devices can generate up to $n^m$ combinations using PP alone. Incorporating DP further expands the strategy space, making real-time scheme selection more complex. In addition, concurrent requests from multiple devices are often processed sequentially, leading to frequent thread switching and reduced scheduling efficiency on the edge server. These runtime bottlenecks significantly limit the scalability and responsiveness of GNN co-inference in practical deployments.

\textbf{Framework-Support Level: Lack of Unified Runtime Scheduling and Communication Support.}
Existing GNN co-inference frameworks lack integrated support for dynamic scheduling and system-aware collaboration. For instance, GCoDE~\cite{zhou2024graph} provides a pipelined engine for static model–partition deployment, but does not support runtime adaptation or hybrid parallelism. Fograph~\cite{zeng2022fograph} uses communication-aware graph partitioning, but it targets specific scenarios and lacks general-purpose scheduling capabilities. Furthermore, the absence of lightweight communication middleware and unified coordination interfaces makes real-time collaboration across multiple edge devices difficult. These limitations constrain the scalability, flexibility, and portability of existing frameworks in real-world edge environments.

\begin{figure*}[t]
    \centering
    \includegraphics[width = 1.0\linewidth]{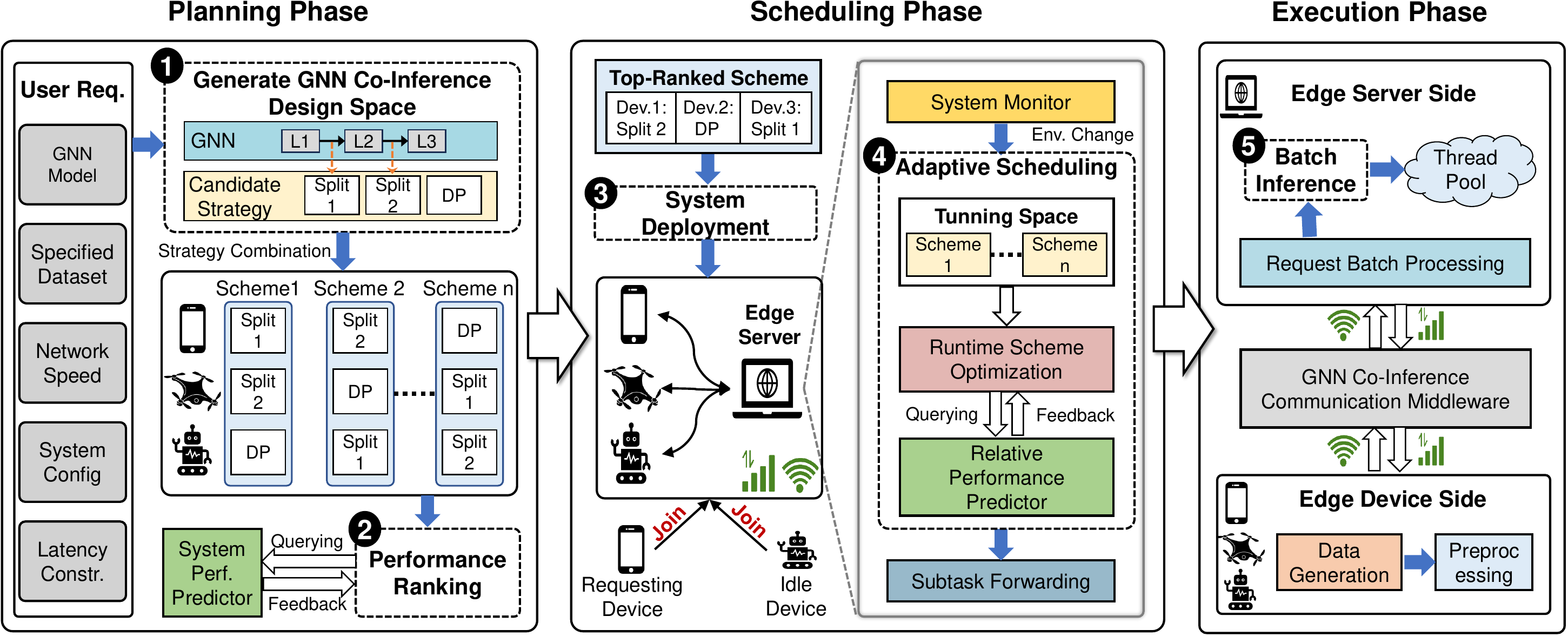}
    \caption{ACE-GNN Overview: A three-phase workflow includes Planning, Scheduling, and Execution Phase.}
    \label{fig:framework}
    \vspace{-6pt}
\end{figure*}

\section{Methodology}\label{sec:methodology}

\subsection{ACE-GNN Overview}
Fig.~\ref{fig:framework} provides an overview of ACE-GNN, which consists of three primary phases: \textit{Planning Phase}, \textit{Scheduling Phase}, and \textit{Execution Phase}. Based on user requirements, ACE-GNN first generates the GNN co-inference design space during the \textit{Planning Phase} (step \raisebox{-0.2em}{\ding[1.3]{182}}). For each edge device, a set of candidate co-inference strategies is derived according to the target GNN model, such as various layer splitting points or DP. The system then enumerates all possible combinations of these per-device strategies across the participating devices. Each combination defines a complete co-inference scheme, specifying the assigned strategy for each device. Leveraging the system performance predictor, ACE-GNN rapidly evaluates and ranks all candidates to identify the optimal scheme (step \raisebox{-0.2em}{\ding[1.3]{183}}). The top-ranked scheme is then deployed to the system (step \raisebox{-0.2em}{\ding[1.3]{184}}). During the \textit{Scheduling Phase}, the system monitor detects changes and triggers adaptive scheduling to maintain performance (step \raisebox{-0.2em}{\ding[1.3]{185}}). In the \textit{Execution Phase}, edge devices preprocess input data and forward it to the edge server for batch inference via tailored communication middleware (step \raisebox{-0.2em}{\ding[1.3]{186}}). The following sections elaborate on the three key components of ACE-GNN: \textbf{System Performance Awareness}, \textbf{Adaptive Scheduling}, and \textbf{Efficient GNN Co-Inference Execution}.

\begin{figure}[t]
    \centering
    \includegraphics[width = 1\linewidth]{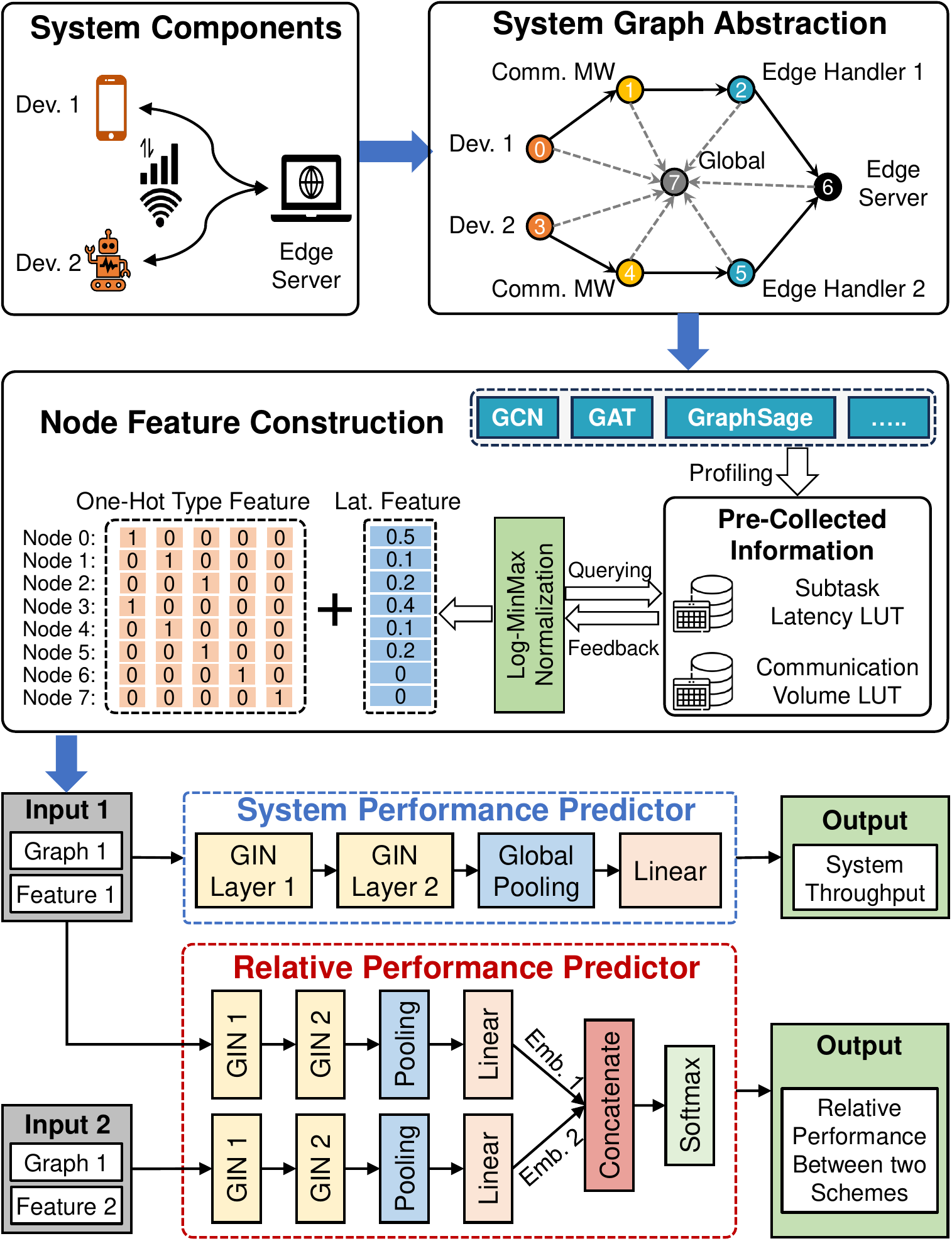}
    \caption{System performance awareness for GNN co-inference in device-edge hierarchies.}
    \label{fig:predictor}
    \vspace{-9pt}
\end{figure}

\subsection{System Performance Awareness}

Achieving adaptive system optimization at runtime necessitates an accurate and efficient system performance awareness method, which remains a major challenge for GNN adaptive co-inference systems and has not yet been effectively addressed.
To this end, ACE-GNN integrates a precise system performance prediction method that effectively estimates the performance of a given collaboration scheme in the current environment and the performance relationship among different schemes.
By leveraging system-level abstraction, ACE-GNN isolates the influence of GNN model structure, application, and system configuration, enabling accurate performance prediction under concurrent multi-device access. Moreover, it achieves high scalability since collecting sub-tasks incurs minimal overhead, allowing rapid adaptation across various models, tasks, and hardware platforms.
Specifically, it includes system graph abstraction, data pre-collection, node feature construction, system performance prediction, and relative performance prediction, as shown in Fig.~\ref{fig:predictor}.
The input to the predictor consists of a system graph derived from the runtime environment and a node feature matrix generated according to the given co-inference scheme. The output includes two performance metrics, defined formally as follows:
\begin{itemize}
    \item \textbf{System Throughput.} Defined as \( \text{Throughput} = \frac{N}{T} \), where \( N \) is the number of completed GNN inference tasks across all devices, and \( T \) is the total execution time. This metric reflects the overall processing capacity of the device-edge system.
    \item \textbf{Relative Performance.} Defined as \( R = \frac{L_2}{L_1} \), where \( L_1 \) and \( L_2 \) are the inference latency under scheme~1 and scheme~2, respectively. A value of \( R > 1 \) indicates that scheme~1 achieves lower latency and better performance.
\end{itemize}

Note that the proposed prediction method is only applied during the scheme optimization, whereas the experiment results in Sec.~\ref{sec:experiment} are directly measured on the target system.

\textbf{System Abstraction.}
Unlike existing methods that focus on model structures, ACE-GNN uses a system-level abstraction to represent the entire edge system as a graph, capturing device properties and their connections.
Specifically, this graph contains nodes for both hardware and software components.
At the hardware level, edge devices and the edge server are represented as nodes, while the communication middleware on each device and the corresponding edge server handlers are represented as software-level nodes.
The edges in this graph represent the data flow within the system, starting at each device node, passing through the middleware to the corresponding edge handler, and ending at the edge server node.
Moreover, global nodes and self-connections are added to enhance message passing during graph learning.
This abstraction minimizes the influence of the GNN structure, considering it only for subtask overhead during node feature construction.
Instead, the device connections and software impacts of the system are considered, greatly enhancing scalability.
Besides, the main difference in predicting various co-inference schemes within the same system graph lies in the initial node features, simplifying system abstraction and enabling a focus on node-level information extraction.

\textbf{Data Pre-Collection.}
In practice, multi-device access introduces significant challenges for prediction, as different devices may deploy different models and applications.
To enable scalable and accurate system performance prediction, ACE-GNN embeds these complex factors into node features and collects execution latency information at the subtask level for each device.
Specifically, ACE-GNN pre-collects subtask latency and communication volume under various GNN models, co-inference schemes, datasets, and devices.
A subtask refers to an inference task where partial GNN layers are assigned to the target device.
These data are stored in lookup tables (LUTs), incurring negligible collection overhead, as GNNs are typically shallow.

\textbf{Node Feature Construction.}
Inspired by \cite{zhou2024graph}, ACE-GNN initializes node features by combining one-hot type encoding with subtask latency information.
The node includes five categories: edge device, communication middleware, edge handler, edge server, and global.
For the latency feature, the edge device and edge handler nodes use the subtask latency of the GNN model under the target scheme on the corresponding device.
In the communication middleware node, latency is estimated based on communication volume and network speed.
Additionally, the latency feature is scaled using the \textit{Log-MinMax} normalization method, as shown below:
\begin{equation}
    x' = \frac{\log(x + 1) - V_{\min}}{V_{\max} - V_{\min}},
\end{equation}
where $V_{\min}$ and $V_{\max}$ are the minimum and maximum values of the log-transformed dataset, respectively.
This normalization process is essential for effective predictor learning (see Sec.~\ref{sec:ablation}).

\textbf{System Performance Prediction.}
Building on the system abstraction, ACE-GNN integrates a GNN-based system performance predictor to estimate system throughput for a given collaborative scheme.
Specifically, the predictor comprises two GIN~\cite{xu2018powerful} layers to capture node information, each with $512$ hidden dimensions.
In addition, the entire system graph information is extracted using a \textit{Global Mean Pooling} layer.
Given the system graph and initialized node features for the specified scheme, the predictor can directly output the system throughput.
In practice, this predictor is primarily used in the \textit{Planning Phase} to quickly determine a deployment scheme based on user-specified system throughput requirements, avoiding the need to examine all candidates.

\textbf{Relative Performance Prediction.}
Accurately predicting the performance of co-inference schemes in multi-device edge systems is challenging, yet the demand for performance stability requires precise runtime evaluations.
Thus, ACE-GNN introduces the relative performance predictor, a clever method derived from the system performance predictor.
It is based on a novel concept: \textbf{runtime system performance awareness aims to assist the scheduling strategy in identifying more efficient solutions, rather than predicting specific values}.
Building on this idea, the complex task of system performance prediction becomes simpler by focusing on comparing the relative performance of schemes, reducing the difficulty of learning.
Specifically, the relative performance predictor includes two GIN feature extractors and a \textit{softmax} for binary classification.
It processes a pair of graphs with the same system architecture, along with distinct initialization features for two collaborative schemes, and outputs their relative performance.
In this manner, predictor learning is simplified and effectively addresses the issue of insufficient training samples by constructing pairs from the collected system performance samples.
The system abstraction and pairwise prediction approach significantly enhance ACE-GNN's scalability, especially compared to the large number of pre-sampled training samples used in existing methods.
Moreover, ACE-GNN uses this predictor to compare the current collaborative scheme with candidate schemes at runtime, ensuring reliable optimization results.
Besides, this predictor is deployed on the edge server, and its runtime overhead is negligible, as the system graph typically contains only a few nodes, requiring approximately $2$ ms per inference.

\subsection{Co-Inference Scheme Planning}
Before deploying the co-inference system, ACE-GNN undergoes a planning phase to design a collaborative scheme in the generated GNN co-inference design space to maximize system performance based on user requirements.
In this phase, ACE-GNN primarily performs system co-inference design space generation and system scheme performance ranking, as detailed below.

\textbf{System Co-Inference Design Space}.
The co-inference design space encompasses a comprehensive set of system collaboration schemes and derives corresponding candidate strategies based on the GNN models designated for deployment on different edge devices.
Specifically, this space contains numerous candidate schemes, each representing a combination of strategies for edge devices to perform co-inference with the edge server.
For example, an edge device running a two-layer GNN model can adopt DP or PP as collaborative strategies with the edge server, where the PP strategy splits the model at the middle layer.

\textbf{System Scheme Performance Ranking}.
In practice, the co-inference scheme design space becomes extremely large due to the diverse GNN models deployed on different edge devices, particularly as the number of connected devices grows.
To improve planning efficiency, ACE-GNN leverages a system performance predictor to rapidly determine a solution that fulfills user requirements.
Specifically, based on the system throughput requirement and iteration limit, ACE-GNN ranks candidates in the design space sequentially until it finds one that meets the throughput requirement or reaches the iteration limit.
After this phase, the chosen system collaborative scheme is generated for later deployment.
Note that \textit{Planning Phase} is an offline procedure that runs once before deployment.

\begin{algorithm}[t]
\KwIn{GNN model $\mathcal{A}$, edge devices set $\mathcal{D}$, edge server $E$, network speed $\mathcal{N}_s$, tuning iteration $T$, preset co-inference strategies: \{$PP_{comp}$, $PP_{comm}$\}.}
\KwOut{ Optimal scheme $\mathcal{P}_{opt} = \{S_1, S_2, \dots, S_n\}$.}
Initialize options $C \leftarrow \{DP, PP_{comp}, PP_{comm}\}$ and $\mathcal{P}_{opt} \leftarrow \emptyset$\hfill \textcolor{brown}{// Select available co-inference strategies}\\
\textbf{/* Stage 1: Coarse-grained Tuning */} \\
Generate tuning space $\mathcal{P} \leftarrow$ All possible schemes for $\mathcal{D}$ from $C$\\ 
\ForEach{$p \in \mathcal{P}$}{
    $\mathcal{P}_{opt} \leftarrow \textnormal{Predictor}(\{\mathcal{P}_{opt}, p\}, \mathcal{D}, E, \mathcal{N}_s, \mathcal{A})$ \hfill \textcolor{brown}{// Rank}\\ 
}
\textbf{/* Stage 2: Fine-grained Tuning */}\\
Initial counter $t = 0$\\
\ForEach{$d_i \in \mathcal{D}$, $S_i \in \mathcal{P}_{opt}$}{ 
    \If{$S_i \neq DP$}{
         ${S_i}' \leftarrow \textnormal{Shift\_left}(S_{i})$\hfill \textcolor{brown}{// Shift split point left}\\ 
        ${S_i}'' \leftarrow \textnormal{Shift\_right}(S_{i})$ \hfill \textcolor{brown}{// Shift split point right}\\
        $\mathcal{P}', \mathcal{P}'' \leftarrow \textnormal{Update}(\mathcal{P}_{opt}, d_i, {S_i}', {S_i}'')$\\
        $\mathcal{P}_{set}\leftarrow \{\mathcal{P}_{opt}, \mathcal{P}', \mathcal{P}''\}$ \hfill \textcolor{brown}{// Update candidates}\\
        $\mathcal{P}_{opt} \leftarrow \textnormal{Predictor}(\mathcal{P}_{set}, \mathcal{D}, E, \mathcal{N}_s, \mathcal{A})$ \hfill \textcolor{brown}{// Rank}\\
    }
\textbf{if} $t > T$ \textbf{then} break \textbf{else} $t \leftarrow t + 1$
}
\textbf{return} $\mathcal{P}_{opt}$ \hfill \textcolor{brown}{// The optimal collaboration scheme}
\caption{Hierarchical scheme optimization.}
\label{algo}

\end{algorithm}
\vspace{-6pt}

\subsection{Adaptive Scheduling}
To achieve runtime optimization, ACE-GNN employs an adaptive scheduling approach based on the relative performance predictor.
Specifically, it comprises two main components: the hierarchical co-inference scheme optimization algorithm and the batch inference strategy.
The former aids in rapidly locating the optimal solution, while the latter boosts processing speed and maximizes resource utilization of edge servers for concurrent multi-device requests.
Moreover, ACE-GNN monitors idle edge devices and forwards subtasks pending at the edge server to them during scheduling to improve system performance.
The scheduling processes are managed by ACE-GNN's communication middleware, which utilizes a specialized event mechanism and task queue for rapid scheme switching.

\textbf{Hierarchical Co-Inference Scheme Optimization.}
Despite the low execution overhead of our predictor, a hierarchical tuning approach is employed to minimize the potential impact of ranking all candidates during runtime.
As shown in Alg.~\ref{algo}, the algorithm inputs include: (1) GNN model $\mathcal{A}$, assuming all edge devices use the same model for clarity; (2) connected edge devices set $\mathcal{D}$; (3) target edge server $E$; (4) network speed limit $\mathcal{N}_s$; (5) fine-grained tuning iteration limit $T$; and (6) preset co-inference strategies: \{$PP_{comp}$, $PP_{comm}$\}.
The preset strategies $PP_{comp}$ and $PP_{comm}$ denote the optimal computational and communication splitting schemes in PP for $\mathcal{A}$, estimated during data pre-collection.
Specifically, $PP_{comp}$ is determined by evaluating all possible splitting points and estimating the total computation cost using a LUT of pre-measured subtask latency.
Likewise, $PP_{comm}$ corresponds to the splitting point that minimizes the intermediate data volume, which can be analytically derived from the model structure.
These preset schemes are stored as efficient initialization options to facilitate runtime optimization.
Given user requirements, ACE-GNN initially configures the set of candidate co-inference strategies $C$ for each device.
Subsequently, a two-stage optimization process is conducted, including coarse-grained and fine-grained tuning, as detailed below.

(1) Coarse-grained tuning.
During coarse-grained tuning, ACE-GNN uses the minimum number of comparisons to determine which collaborative mechanism, DP or PP, is more suitable for the current environment.
In this stage, ACE-GNN first generates a tuning space $\mathcal{P}$ that includes all schemes, where each scheme consists of a co-inference method chosen from available options $C$ for each device in $\mathcal{D}$.
The designed relative performance predictor is then used to rank all the schemes in $\mathcal{P}$ and identify the optimal plan $\mathcal{P}_{opt}$, which includes a suitable co-inference scheme for each edge device.
Following that stage, ACE-GNN will perform additional tuning if $\mathcal{P}_{opt}$ contains PP schemes.

(2) Fine-grained tuning.
During the fine-grained tuning, potential splitting schemes $S_i'$ and $S_i''$ are created by shifting the split points of the current scheme $S_i$ for $d_i$ to the left and right.
The final optimal plan $\mathcal{P}_{opt}$ is identified after repeatedly comparing the current optimal scheme with the updated plans $\mathcal{P}'$ and $\mathcal{P}''$ until the specified limit $T$ is reached.
In practice, edge devices with similar hardware characteristics can use the same scheme to reduce tuning overhead.
Note that within the tolerable overhead, this stage can further expand the scheme space by iteratively shifting the splitting points toward more distant neighboring points.
Moreover, the final optimal scheme $\mathcal{P}_{opt}$ denotes the best-performing strategy selected from the candidate space using the performance predictor.

\begin{figure}[t]

    \centering
    \includegraphics[width = 1\linewidth]{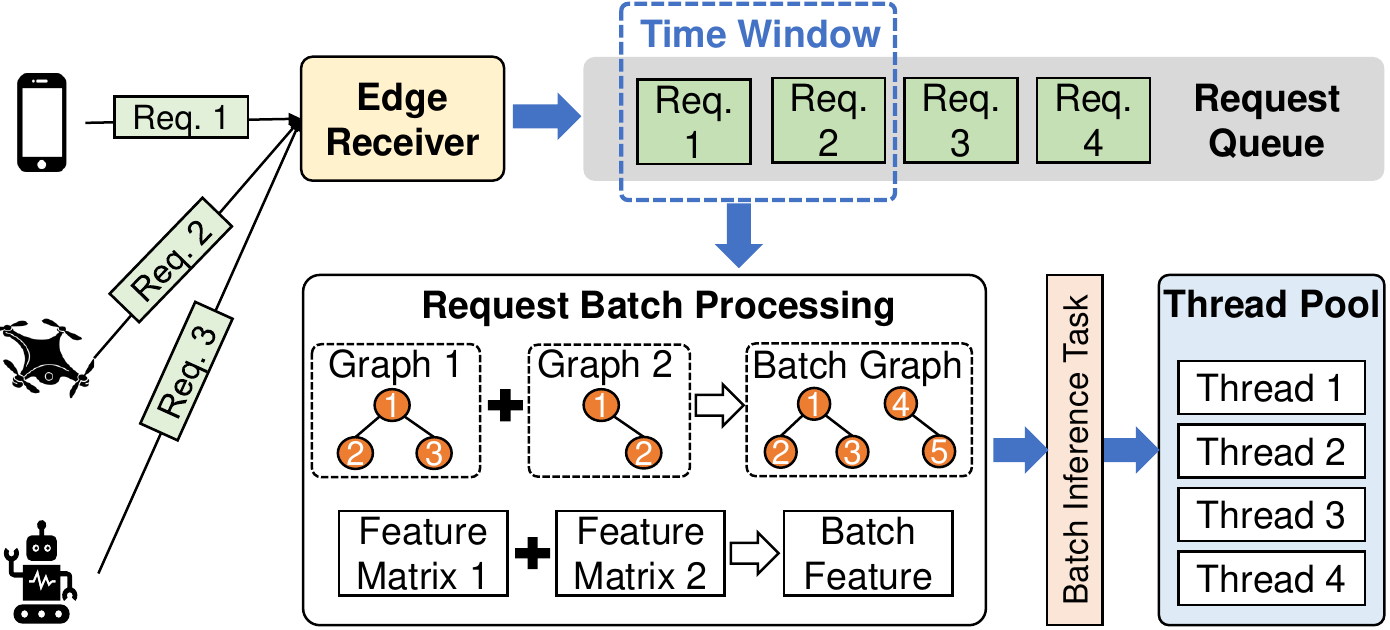}
    \caption{Batch processing of requests from multiple devices.}
    \label{fig:batch_inference}
    \vspace{-6pt}
\end{figure}

\textbf{Batch Inference Strategy.}
To enhance the efficiency of edge server request processing, ACE-GNN employs a batch inference strategy.
As depicted in Fig.~\ref{fig:batch_inference}, a co-inference request from an edge device initially enters the request queue at the edge server to await processing.
In the request queue, ACE-GNN sets a time window that triggers batch processing once it exceeds a predefined limit or the number of requests surpasses the maximum batch size.
During processing, this batch of requests is combined, incorporating the graph data and feature information from each request, forming a batch inference task.
The task is then submitted to an idle thread in the thread pool for specific inference execution.
Finally, the inference result is divided and sent back to each corresponding edge device according to the request information recorded in the queue.
Through batch processing requests, the edge server’s efficiency is significantly improved, especially for GPUs with powerful parallel processing capabilities.
In practice, larger batch sizes are not always better, and an appropriate batch size limit is needed to fully utilize edge server performance (see Sec.~\ref{sec:ablation}).

\subsection{Efficient GNN Co-Inference Execution}

Due to the absence of specialized communication libraries for GNN co-inference, ACE-GNN implements an efficient communication middleware based on \texttt{Python asyncio}~\cite{asyncio}.
The communication process within the device-edge hierarchy is managed through asynchronous events, enabling maximal parallelization of communication and computation.
Moreover, an efficient GNN co-inference engine is developed based on the PyTorch Geometric (PyG) framework~\cite{fey2019fast} to support batch task processing and task scheduling.
Leveraging these specialized designs, ACE-GNN is able to rapidly adapt to system changes at runtime.

\textbf{Communication Middleware.}
The communication middleware consists of device-side and edge-side components, with data organized as message packets and exchanged between devices and the edge server.
Additionally, each communication packet includes a customized message header to support scalable scheduling, which includes three parts: message type, task ID, and message size.
Specifically, there are three types of messages: \textit{Scheduling}, \textit{Task}, and \textit{Result}.
Among them, \textit{Scheduling} messages represent control signals such as start, pause, and scheme update; \textit{Task} messages carry data for co-inference; and \textit{Result} messages deliver the final inference result.
Furthermore, all packets are compressed and serialized to minimize overhead.

\textbf{Inference Engine.}
ACE-GNN leverages thread pool technology on edge servers to enhance parallel processing.
Moreover, all devices use \texttt{asyncio} queues to store task information and ensure quick task access.
Each task block contains complete task information, including ID, device type, source device IP, task type, task data, arrival time, deployed model, and co-inference scheme.
Based on this, each task block is processed independently, ensuring that adaptive scheduling does not lose any received task information.
Besides, the inference engine includes single inference and batch inference modules, and employs non-blocking task processing to further reduce computational waiting overhead.

\begin{figure}[t]

    \centering
    \includegraphics[width = 1\linewidth]{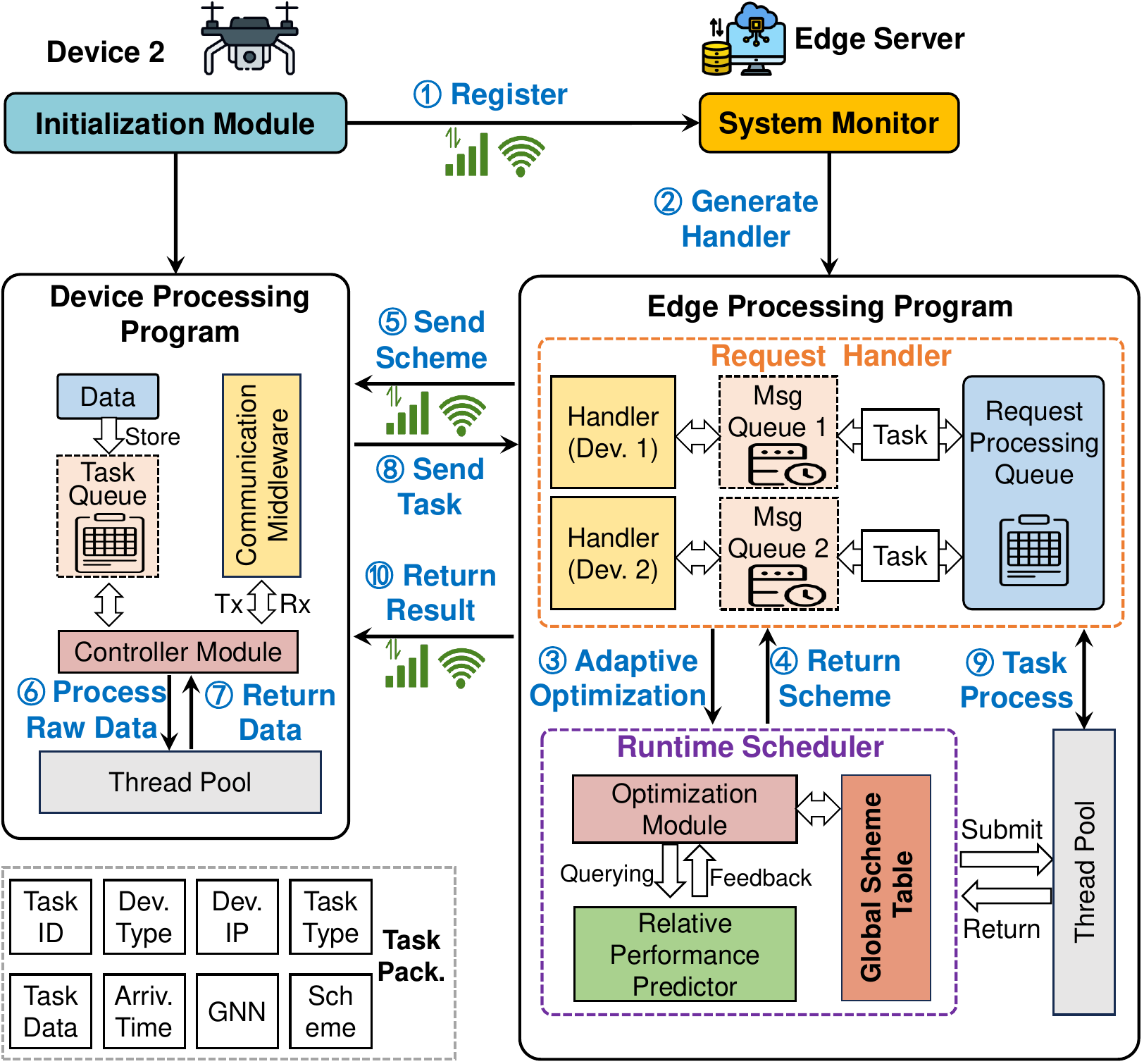}
    \caption{New edge device access processing flow.}
    \vspace{-9pt}
    \label{fig:engine}
\end{figure}

\textbf{New Device Connection Workflow.} Fig.~\ref{fig:engine} shows the workflow for handling a new device connection. When a new edge device (e.g., Device 2) joins the system, it sends a registration request (step~\circnum{1}), which is processed by the \textit{System Monitor} on the edge server. Upon successful registration, the system creates a dedicated handler for the device (step~\circnum{2}) within the \textit{Edge Processing Program} to support asynchronous communication. The \textit{Runtime Scheduler} is then activated to perform adaptive optimization (step~\circnum{3}), selecting a collaborative scheme based on the current system state. This selection is guided by the \textit{Relative Performance Predictor}. The chosen scheme is returned to the device (steps~\circnum{4} and \circnum{5}). After receiving the scheme, the edge device updates its local controller, preprocesses data, and prepares inference tasks (steps~\circnum{6} and \circnum{7}). These tasks are sent to the edge server (step~\circnum{8}), placed into a message queue, and dispatched to the \textit{Request Processing Queue} for execution by the server’s thread pool (step~\circnum{9}). Finally, the inference result is sent back to the edge device (step~\circnum{10}).

Meanwhile, the edge server notifies other edge devices that require switching based on the optimized system scheme.
Note that the adaptive optimization process runs only on idle threads of the edge server to prevent blocking other collaborative tasks.
During this process, the edge device only needs to receive and update the subsequent processing scheme, with minimal overhead enabled by our specialized communication middleware.
Moreover, to reduce the overhead of frequent scheme changes, the edge server triggers adaptive scheduling only when system changes surpass specified thresholds, such as device access or significant network degradation.
In practice, the overhead of adaptive optimization is typically within milliseconds, with minimal impact on performance compared to system changes.

\section{Experiment}\label{sec:experiment}
\subsection{Experimental Settings} \label{sec:setting}

\textbf{Models, Datasets and Competitor Settings.}
Our evaluation considers five GNN models, including widely used GCN~\cite{kipf2016semi}, GAT~\cite{velivckovic2018graph}, GraphSAGE~\cite{hamilton2017inductive}, DGCNN~\cite{wang2019dynamic}, and optimal co-inference models designed by GCoDE~\cite{zhou2024graph}.
For GCoDE and PAS, the model are implemented following the structure reported in their original papers, while other typical GNNs are implemented using instances from the PyTorch Geometric (PyG) model zoo~\cite{fey2019fast}.
We use the point cloud dataset ModelNet40~\cite{wu20153d} with 1024 points and MR text dataset~\cite{zhang2020every}, following the evaluation settings in~\cite{zhou2024graph}. 
Additionally, the IoT dataset SIoT~\cite{marche2020exploit} and Yelp social network dataset~\cite{rayana2015collective} are used, adhering to the processing method in~\cite{zeng2022fograph}.
To benchmark ACE-GNN with SOTA works, we consider five competitors: (1) the GNN co-inference framework GCoDE~\cite{zhou2024graph}, (2) the co-inference method Branchy~\cite{shao2021branchy}, (3) GNN hardware-aware architecture search framework HGNAS~\cite{zhou2023hardware}, (4) GNN architecture search method PAS~\cite{wei2023neural}, and (5) the GNN distributed inference framework Fograph for edge scenarios~\cite{zeng2022fograph}.
Furthermore, we considered standard performance metrics for edge scenarios, including inference latency, system throughput, and on-device energy consumption.

\textbf{Edge Environment and Implementation Settings.}
To evaluate ACE-GNN, we consider various system configurations, including four edge device types: Jetson TX2~\cite{tx2}, Jetson Nano~\cite{nano}, Raspberry Pi4B~\cite{pi4b}, Raspberry Pi3B~\cite{pi3b}, and two edge servers: Nvidia 1060 GPU~\cite{1060} and Intel i7-7700 CPU~\cite{cpu}.
All devices are connected to a wireless router, and network conditions are varied by setting bandwidth limits using the Linux \texttt{tc} tool.
To train the predictor, we collect 2000 co-inference samples covering various systems, collaboration schemes, datasets, and GNN models (70\% for training and 30\% for validation).
MAPE and BCE serve as the loss functions for the system and relative performance predictors, respectively.
During evaluation, single-device access was evaluated using the system performance predictor, whereas the relative performance predictor was used in multi-device scenarios.
For batch inference, the maximum size and time window are set to $5$ and $10$ ms, respectively.
All implementations and experiments are based on PyG.

\begin{figure}[t]
    \centering
    \includegraphics[width = 1\linewidth]{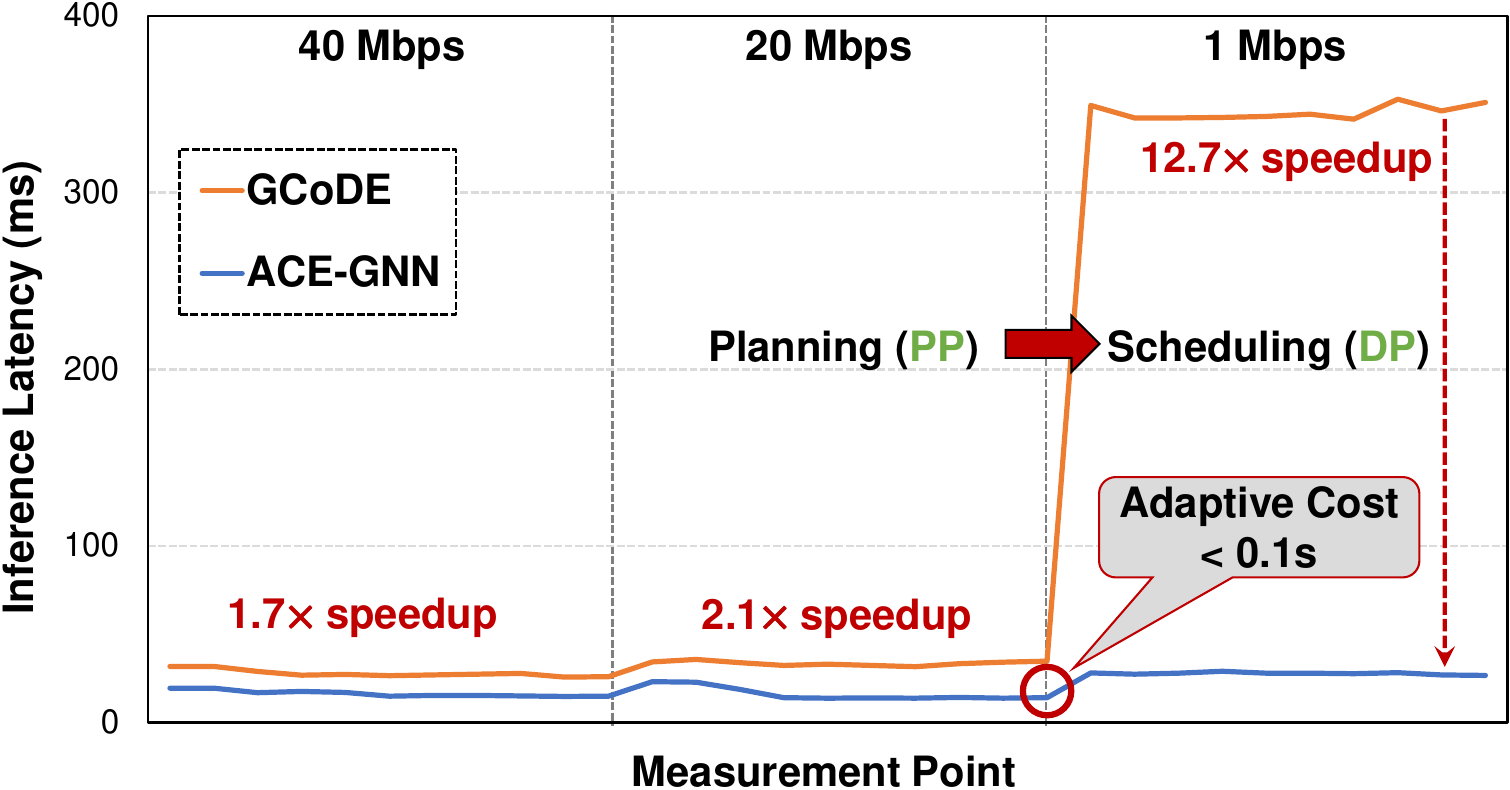}
    \caption{Device-edge co-inference latency variation under network deterioration.}
    \label{fig:net_change_compare}
\end{figure}

\begin{table}[t]
\renewcommand\arraystretch{1.3}
\centering
\caption{Inference latency comparison under varying network speeds on ModelNet40.}
\label{tab:mn_compare}
\resizebox{1\linewidth}{!}{%
\begin{tabular}{|c|c|c|c|c|c|}
\hline
 \parbox[c][0.8cm]{0.9cm}{\centering \textbf{$\mathcal{N}_s$\\(Mbps)}}  &
 \textbf{Method} & 
 \parbox[c]{1.45cm}{\centering \textbf{TX2 - GPU \\ (ms)}} &
  \parbox[c]{1.45cm}{\centering \textbf{TX2 - CPU \\ (ms)}} &
  \parbox[c]{1.4cm}{\centering \textbf{Pi - GPU \\ (ms)}} & 
  \parbox[c]{1.4cm}{\centering \textbf{Pi - CPU \\ (ms)}} \\ \hline
\multirow{4}{*}{\textbf{100}} & HGNAS            & 52.1  & 52.1  & 241.5 & 241.5 \\
                    & Branchy          & 141.4 & 138.9  & 530.9  & 526.1  \\
                    & GCoDE            & 30.4   & 26.1  & 20.4   & 38.6  \\
                    & \textbf{ACE-GNN} & \textbf{13.7}  & \textbf{12.7}  & \textbf{8.3}   & \textbf{28.4}  \\ \hline
\multirow{4}{*}{\textbf{40}} & HGNAS            & 52.1  & 52.1  & 241.5 & 241.5 \\
                    & Branchy          & 141.2 & 140.2 & 541.8 & 528.1 \\
                    & GCoDE            & 31.9  & 21.0  & 25.0  & 64.4  \\
                    & \textbf{ACE-GNN} & \textbf{13.9}  & \textbf{14.0}  & \textbf{8.3}   & \textbf{29.2}  \\ \hline
\multirow{4}{*}{\textbf{20}} & HGNAS            & 52.1  & 52.1  & 241.5 & 241.5 \\
                    & Branchy          & 141.0 & 140.4 & 555.5 & 560.1 \\
                    & GCoDE            & 34.9  & 31.2  & 30.5  & 43.1  \\
                    & \textbf{ACE-GNN} & \textbf{14.1}  & \textbf{14.0}  & \textbf{8.2}   & \textbf{29.0}  \\ \hline
\multirow{4}{*}{\textbf{1}}  & HGNAS            & 52.1  & 52.1  & 241.5 & 241.5 \\
                    & Branchy          & 142.0 & 141.0 & 557.4 & 545.8 \\
                    & GCoDE            & 145.8 & 343.1 & 278.0 & 141.6 \\
                    & \textbf{ACE-GNN} & \textbf{33.5}  & \textbf{26.9}  & \textbf{63.1}  & \textbf{57.1}  \\ \hline
\end{tabular}%
}

\end{table}

\subsection{Evaluation under Network Variability}

Fig.~\ref{fig:net_change_compare} compares ACE-GNN and the SOTA co-inference framework GCoDE on ModelNet40, using GCoDE's model with Jetson TX2 and Intel CPU as the device and edge.
This comparison highlights that GCoDE suffers significant performance degradation as the network deteriorates, due to its lack of runtime scheduling and reliance on PP.
In contrast, ACE-GNN adaptively schedules from PP to DP during network changes, reducing communication overhead and ensuring stable runtime performance, resulting in a $12.7\times$ speedup over GCoDE.
Such performance gains align with the characteristics of PP and DP as analyzed in \textbf{Key Insight}.
Due to the integration of system performance predictor and dedicated communication middleware, adaptive scheduling introduces negligible overhead (under $0.1$s).

Detailed experimental results on ModelNet40 are shown in 
Tab.~\ref{tab:mn_compare}, where ACE-GNN outperforms all competitors under varying network conditions.
Under better network conditions ($\mathcal{N}_s \leq 40\,\text{Mbps}$), ACE-GNN’s execution engine and communication middleware effectively balance computation and communication overhead, achieving up to $29.1\times$, $65.2\times$, and $3.0\times$ speedup over HGNAS, Branchy, and GCoDE, respectively.
Even when network speed drops below 1 Mbps, ACE-GNN sustains high performance through adaptive scheduling and data parallelism, reaching speedups of $4.2\times$, $9.5\times$, and $12.7\times$ over HGNAS, Branchy, and GCoDE, respectively.
In addition, when the bandwidth is increased to $100$ Mbps, performance improvements across all methods become marginal, as the system reaches a computational bottleneck where network overhead no longer dominates execution time.
Besides, ACE-GNN shows greater performance gains under device-edge configurations with pronounced heterogeneity, such as Jetson TX2 with an Intel CPU or Raspberry Pi4B with an NVIDIA GPU.
This is attributed to the system performance predictor’s effective awareness of network conditions and hardware heterogeneity, enabling adaptive scheduling of appropriate collaboration strategies.

\begin{figure}[t]
    \centering
    \includegraphics[width = 1\linewidth]{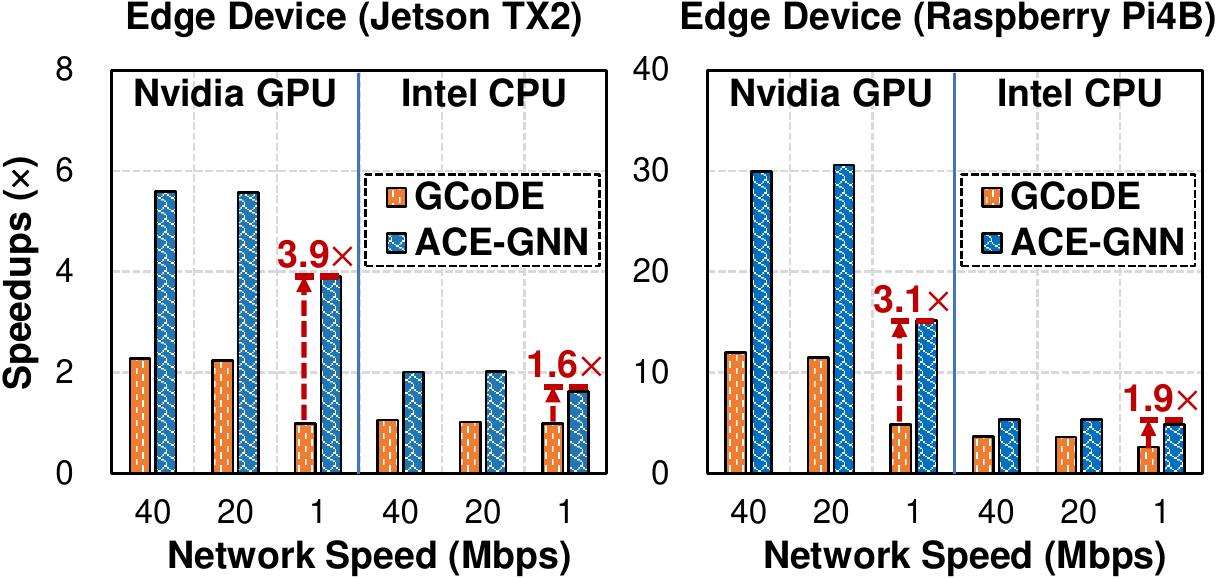}
    \caption{Comparison of DGCNN co-inference speedup on ModelNet40.}
    \label{fig:dgcnn}
    \vspace{-9pt}
\end{figure}

Fig.~\ref{fig:dgcnn} illustrates the performance gains of co-inference when deploying a larger DGCNN model, using direct on-device inference as the baseline.
It can be seen that ACE-GNN consistently delivers significant performance gains across diverse network conditions and device-edge setups.
As Raspberry Pi4B with weaker computing power is employed as the edge device, ACE-GNN can achieve up to $30.6\times$ acceleration over on-device inference under $40$ Mbps, and maintains up to $15.2\times$ even with a $1$ Mbps network limit, highlighting its strength in resource-constraint environments.
Compared to GCoDE, ACE-GNN achieves up to $3.9\times$ and $3.1\times$ speedup on Jetson TX2 and Raspberry Pi4B, respectively, despite both adopting a device-edge co-inference paradigm.
This performance advantage is due to the architectural characteristics of DGCNN, where the intermediate feature representations have high dimensionality. In such cases, GCoDE adopts a PP strategy that introduces significant communication overhead when partitioning the model across devices, thereby limiting the benefits of collaboration. In contrast, ACE-GNN recognizes this property and adaptively selects the DP strategy to minimize communication volume, resulting in more efficient inference.

\begin{figure}[t]
    \centering
    \includegraphics[width = 1\linewidth]{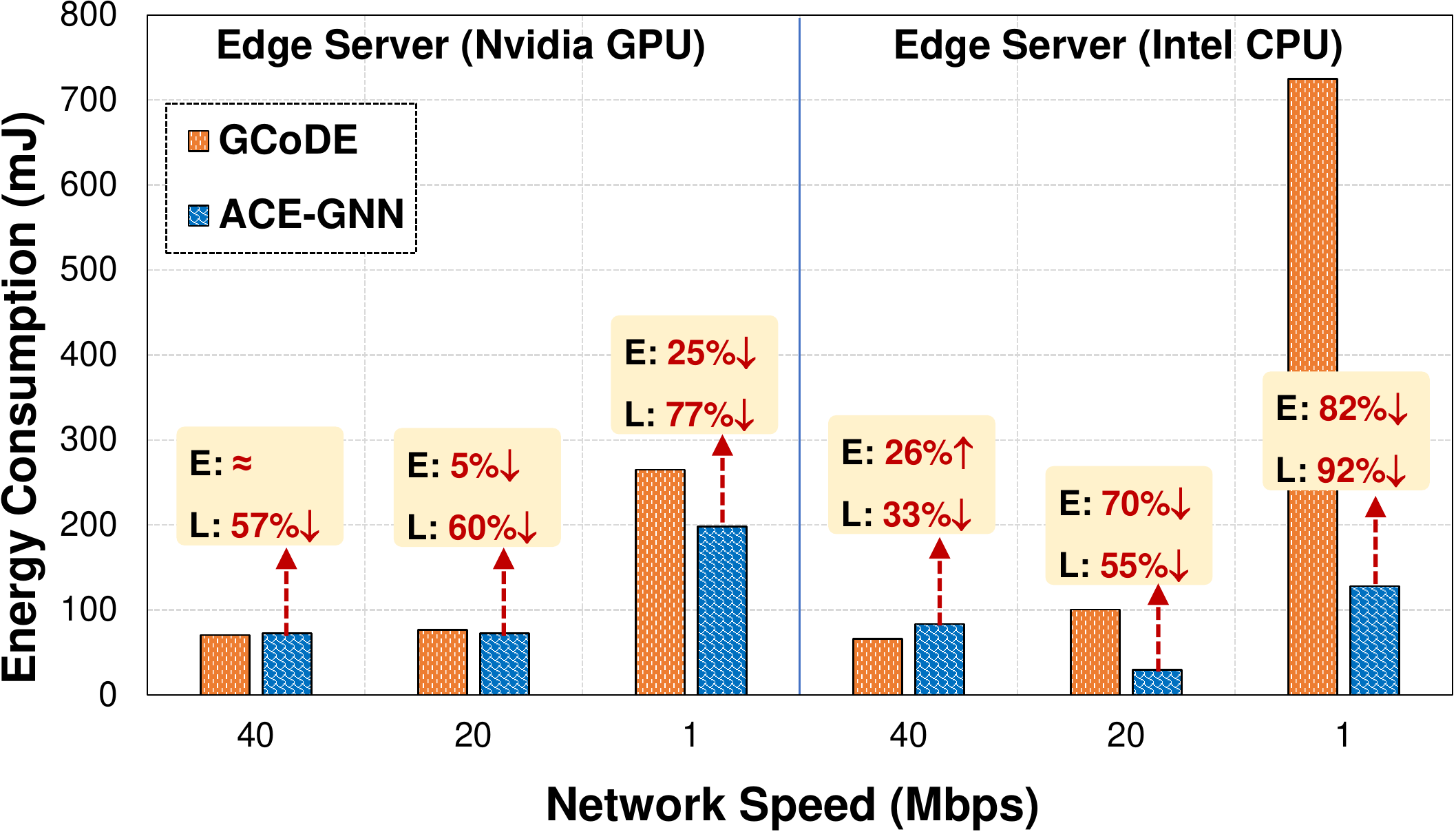}
    \caption{Evaluation of on-device energy consumption durning co-inference on Jetson TX2.}
    \label{fig:energy}
    \vspace{-6pt}
\end{figure}

To evaluate energy consumption, we perform on-device measurements on Jetson TX2 using its on-board power sensors. The Jtop tool is used to record average runtime power during co-inference, and energy per inference is estimated by multiplying power with execution time. Fig.~\ref{fig:energy} shows a comparison of ACE-GNN and GCoDE under different network conditions. With sufficient bandwidth, ACE-GNN reduces latency by up to $57\%$ with similar energy usage. Under constrained bandwidth, it significantly reduces both latency and energy. For instance, ACE-GNN achieves $25\%$ energy and $77\%$ latency reduction on an Nvidia GPU server, and $82.3\%$ energy and $92\%$ latency reduction on the Intel CPU server. These results confirm that ACE-GNN can adapt to diverse environments by selecting efficient schemes, improving resource utilization and energy efficiency.

\begin{figure}[t]
    \centering
    \includegraphics[width = 1\linewidth]{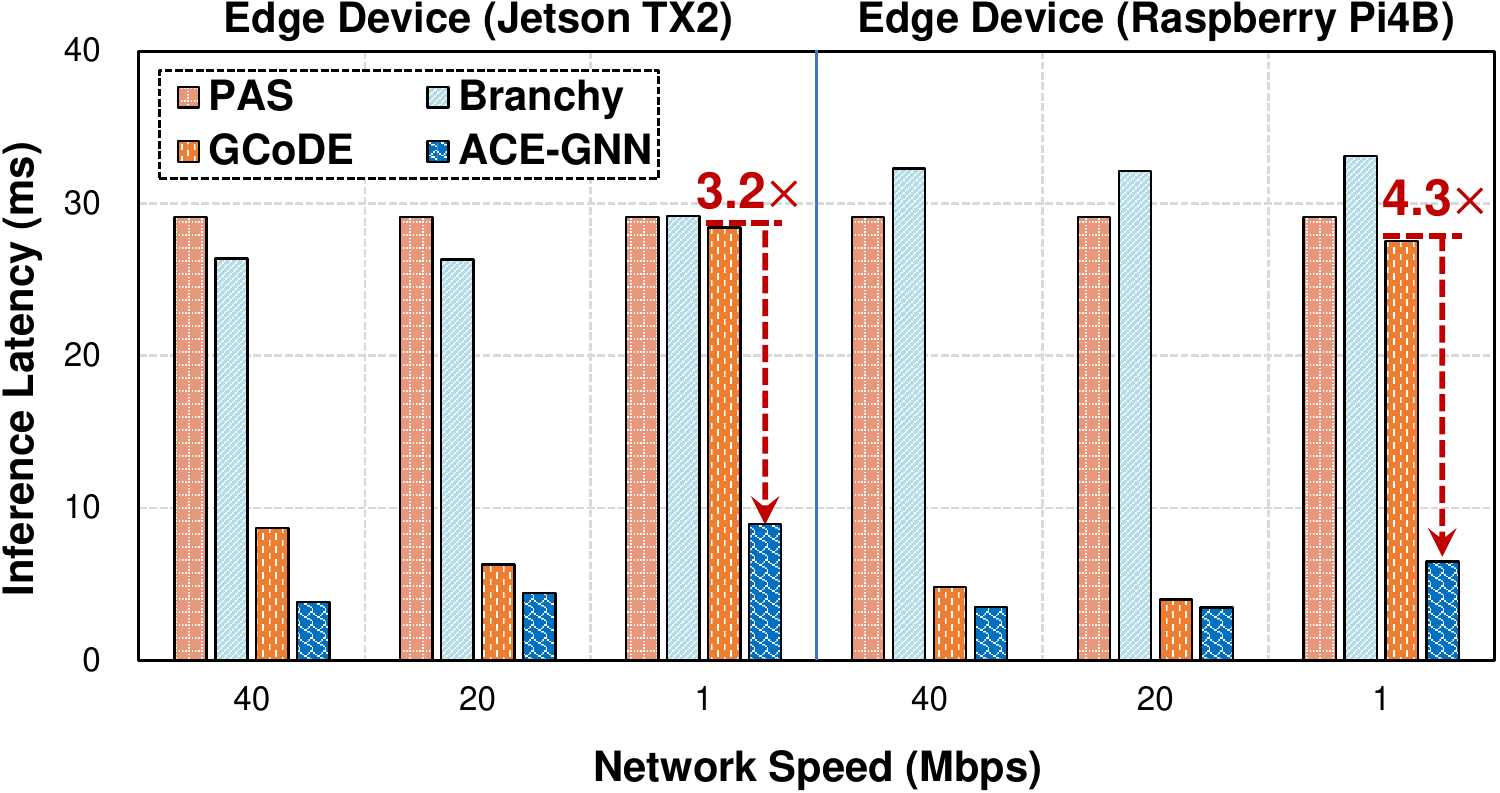}
    \caption{Performance comparison under varying network conditions on the MR dataset, with an Nvidia GPU as the edge server.}
    \label{fig:mr}
    \vspace{-9pt}
\end{figure}

Fig.~\ref{fig:mr} presents the experimental results on the MR dataset under different network conditions, comparing ACE-GNN with PAS, Branchy, and GCoDE. Unlike the ModelNet40 dataset, MR exhibits an opposite data profile, characterized by significantly fewer points ($1024$ vs. $17$) and much higher node feature dimensionality ($3$ vs. $300$). These properties lead to different collaboration preferences. At $40$ Mbps, ACE-GNN leverages the DP to improve device-edge resource utilization by offloading a larger portion of computation. In contrast, at $1$ Mbps, it adaptively switches to the PP, allowing task partitioning at an earlier stage where intermediate features have lower dimensions, thus effectively reducing communication overhead.
As a result, ACE-GNN achieves up to $7.5\times$, $9.2\times$, and $2.2\times$ speedup over PAS, Branchy, and GCoDE, respectively, at 40 Mbps, and $3.2\times$, $5.1\times$, and $4.3\times$ at 1 Mbps.

\begin{figure}[t]
    \centering
    \includegraphics[width = 1\linewidth]{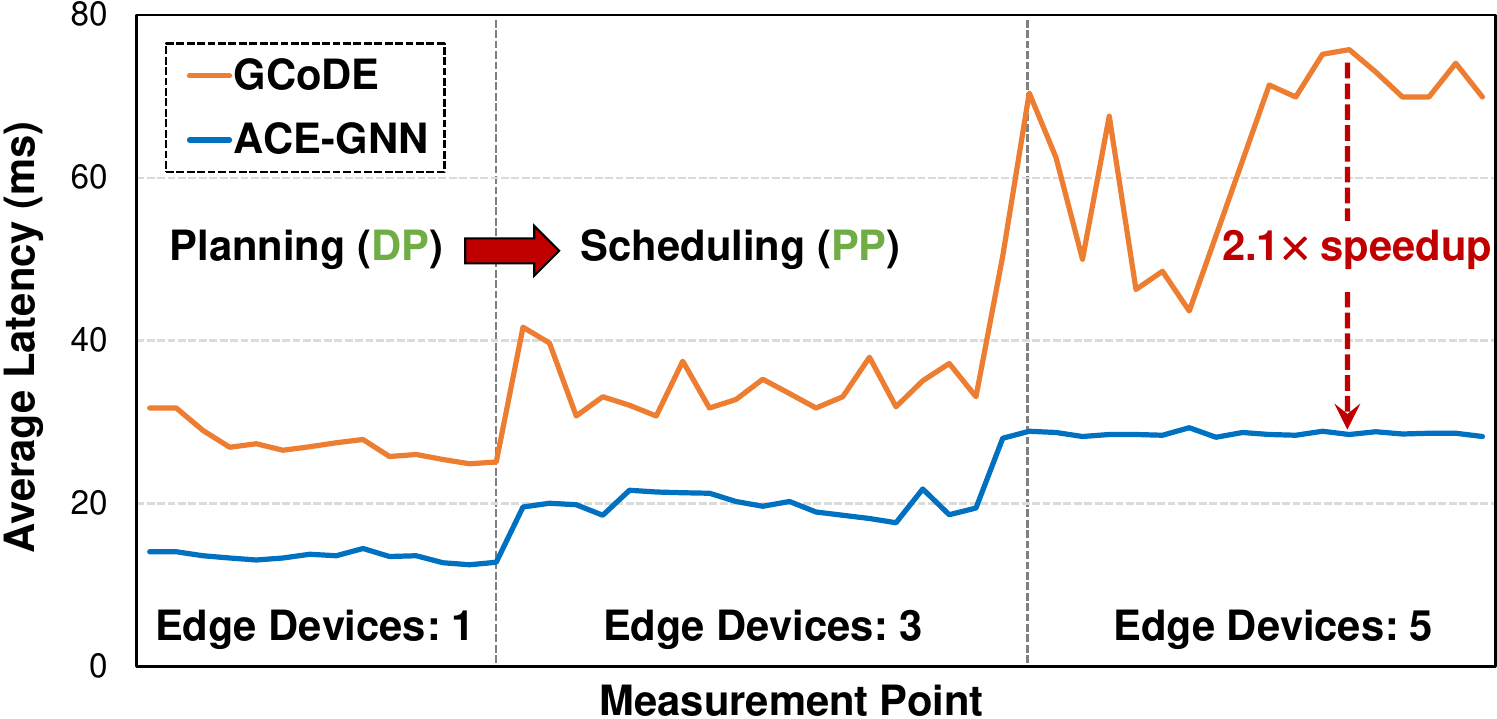}
    \caption{System performance variation with multi-device access.}
    \label{fig:device_change_compare}
\end{figure}

\begin{figure}[t]
    \centering
    \includegraphics[width = 1\linewidth]{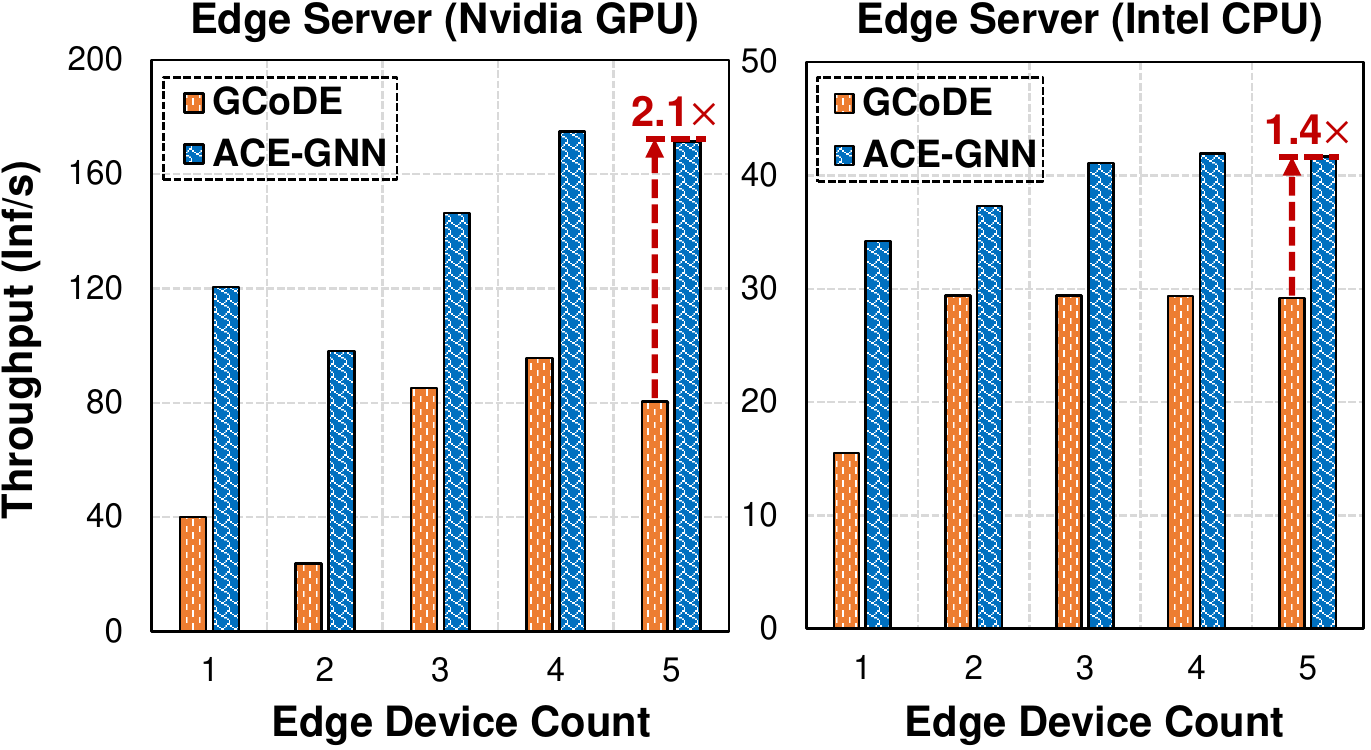}
    \caption{Comparison of throughput in the edge systems with multiple devices.}
    \label{fig:multi_device}
    \vspace{-6pt}
\end{figure}

\subsection{Evaluation under Edge Workload Variability}

To evaluate ACE-GNN under heavy edge workloads, we compare it with GCoDE across varying numbers of edge devices on ModelNet40, each making independent co-inference requests.
The network speed is set to $40$ Mbps, and all edge devices use Raspberry Pi4B to minimize interference.
As shown in Fig.~\ref{fig:device_change_compare}, with Nvidia GPU as the edge server, GCoDE shows considerable variation in efficiency as more edge devices access the system.
In contrast, ACE-GNN reduces edge workload and maintains stable performance through system-aware scheduling, resulting in up to $2.1\times$ speedup over GCoDE.
It dynamically switches from the DP to the PP scheme with an optimal split point to exploit system heterogeneity.
Additionally, the batch inference strategy enables ACE-GNN to handle device requests in parallel, fully utilizing GPUs' parallel capabilities.

Detailed multi-device access evaluation results are shown in Fig.~\ref{fig:multi_device}.
When using an Nvidia GPU as the edge server, ACE-GNN improves system throughput by $4.1\times$ with two devices and maintains a $2.1\times$ gain with five devices, compared to GCoDE, by leveraging adaptive scheduling and batch processing.
Additionally, ACE-GNN achieves a $1.4\times$ speedup over GCoDE when using an Intel CPU as the edge server.
With hierarchical optimization and efficient predictors ($2$ ms per inference), ACE-GNN's adaptive scheduling overhead is minimal and in the millisecond range.

\begin{figure}[t]
    \centering
    \includegraphics[width = 1\linewidth]{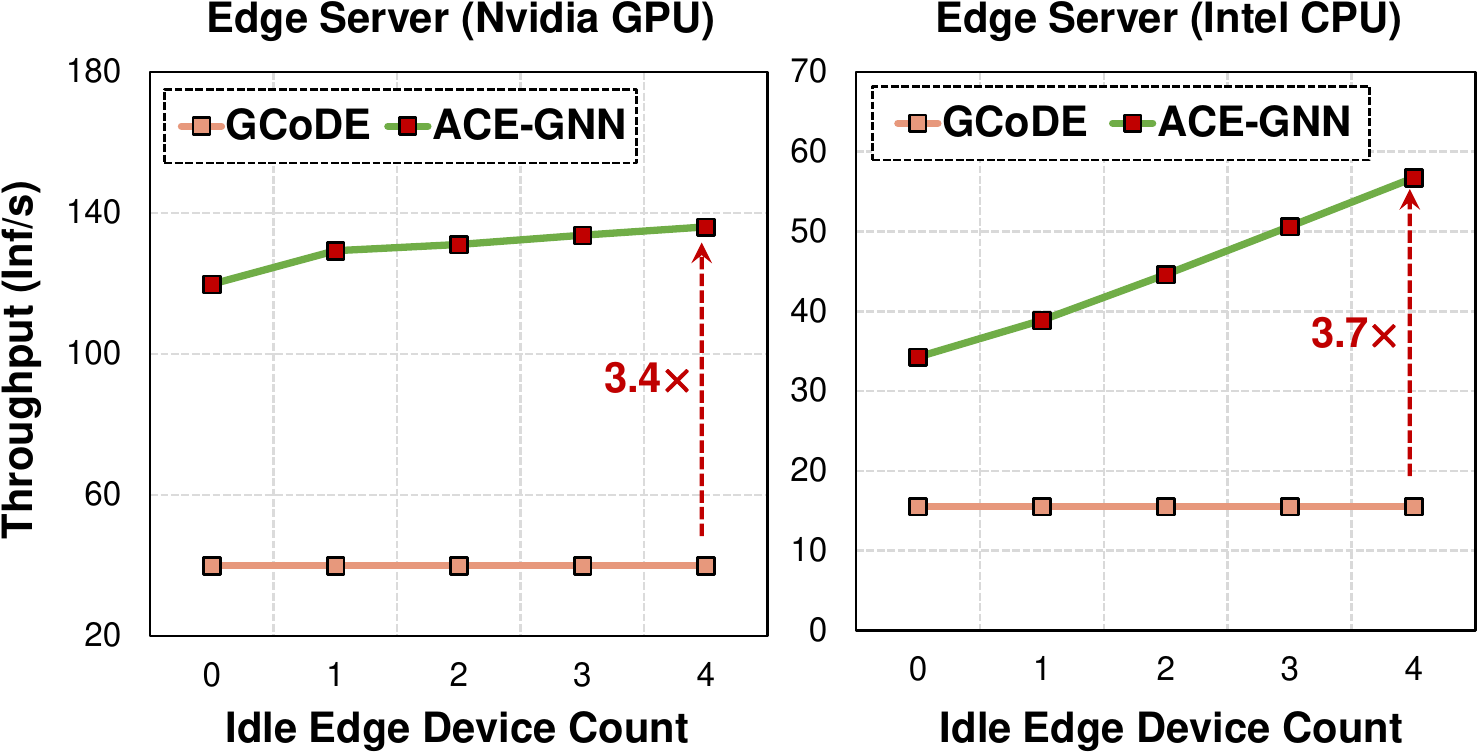}
    \caption{System performance improvement with available idle edge devices.}
    \label{fig:available}
\end{figure}

\begin{figure}[t]
    \centering
    \includegraphics[width = 1\linewidth]{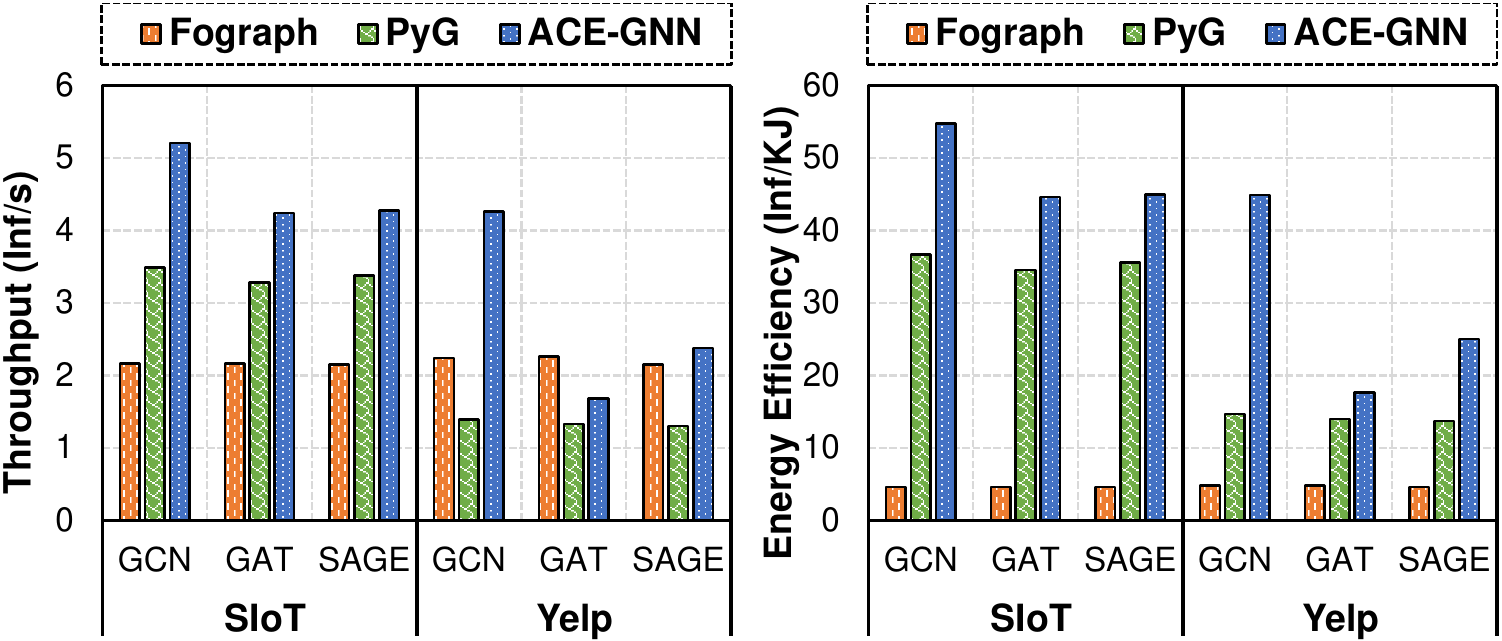}
    \caption{System performance comparison on larger SIoT and Yelp dataset.}
    \label{fig:s3_yelp}
\end{figure}

\subsection{Evaluation with Available Idle Edge Devices}

As shown in Fig.~\ref{fig:available}, ACE-GNN forwards subtasks to idle devices through adaptive scheduling, achieving $3.4\times$ and $3.7\times$ higher throughput than GCoDE.
Such improvement is more significant for the Intel CPU, as its relatively lower computational power amplifies the role of idle resource utilization.
Additionally, Fig.~\ref{fig:s3_yelp} compares ACE-GNN with Fograph and PyG on the larger SIoT ($16216$ nodes, $52$ features) and Yelp ($10000$ nodes, $100$ features) datasets.
Fograph uses six Intel CPUs, while ACE-GNN only uses four idle Raspberry Pi4B as edge devices and one Intel CPU as the edge server.
PyG is deployed in the same hardware environment as ACE-GNN for a fair comparison.
It is clear that ACE-GNN outperforms Fograph across various GNN models and datasets, achieving up to $2.4\times$ higher throughput and $11.7\times$ better energy efficiency.
While ACE-GNN's throughput is slightly lower for GAT due to device limitations, it still achieves significant energy efficiency gains.
These gains are due to accurate prediction and adaptive scheduling, enabling ACE-GNN to select the optimal co-inference scheme and fully utilize resources.
For the SIoT dataset, ACE-GNN uses the DP and switches to PP for the Yelp dataset, except for GAT, to reduce communication overhead, aligning with the analysis in \textbf{Key Insight}.
Moreover, compared to the distributed execution in PyG, ACE-GNN still achieves a $3\times$ efficiency improvement, demonstrating its superior performance.

\begin{figure}[t]
    \centering
    \includegraphics[width = 1\linewidth]{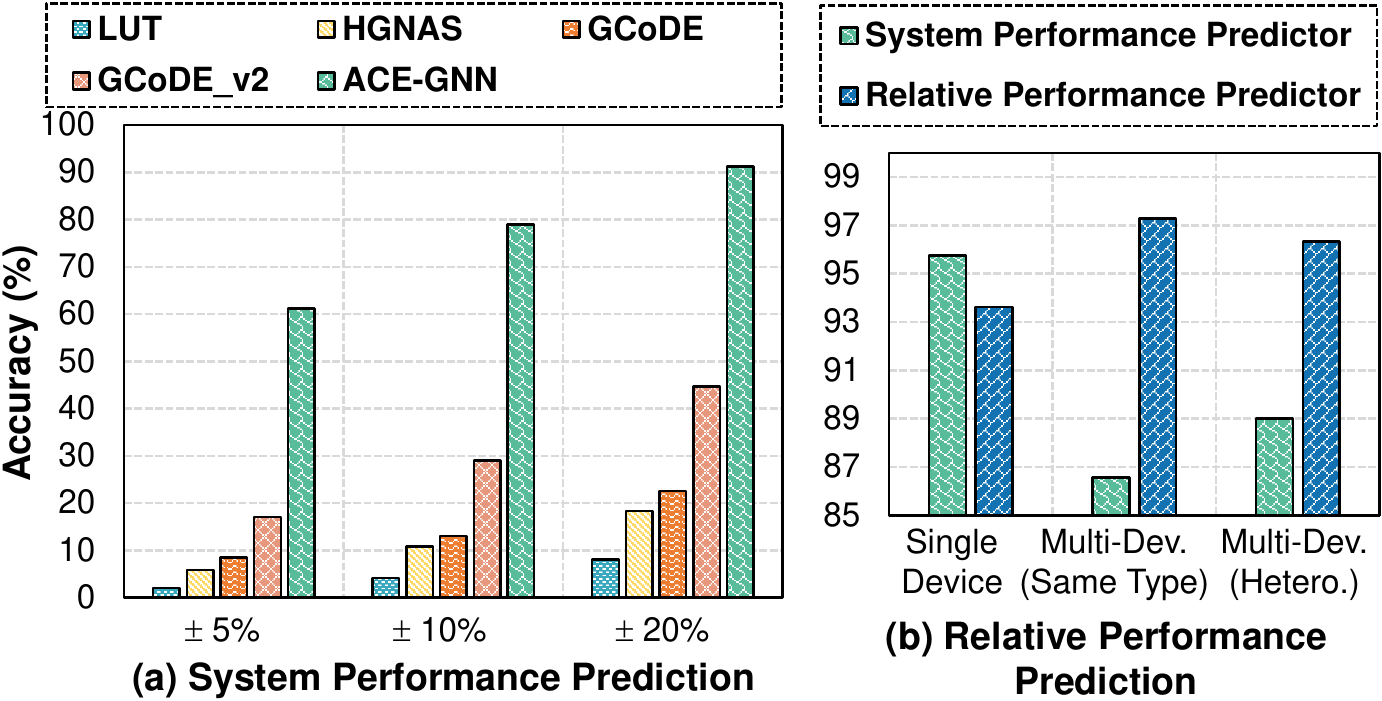}
    \caption{System performance prediction and relative performance prediction accuracy in complex device-edge hierarchies with multi-device access.}
    \label{fig:predictor_result}
\end{figure}

\subsection{Evaluation of System Performance Awareness Accuracy}

Fig.~\ref{fig:predictor_result} (a) shows that ACE-GNN achieves prediction accuracy of about $80\%$ and $91\%$ within $10\%$ and $20\%$ error bounds, respectively, outperforming all baselines.
While GCoDE considers heterogeneous and network factors, it neglects the impact of multi-device access and focuses on model structure, necessitating a large training sample for learning.
Although GCoDE\_v2 improves with our normalization and training strategy, its accuracy remains less than half.
In contrast, ACE-GNN's system-level abstraction effectively handles such problems, achieving optimal accuracy.
Furthermore, Fig.~\ref{fig:predictor_result} (b) shows the results of predicting relative performance between collaborative schemes.
In this context, ACE-GNN’s system performance predictor achieves up to $95.8\%$ accuracy with a single device but performs poorly with multi-device access.
Conversely, the relative performance predictor achieves up to $97.3\%$ accuracy, ensuring the effectiveness of runtime scheme optimization.
This is attributed to the simplification of the system-awareness problem and the larger training samples generated by constructing sample pairs for the relative performance predictor.

\subsection{Scalability and Extensibility Evaluation}\label{sec:scalability}
To demonstrate the practical effectiveness of ACE-GNN, we perform experiments evaluating its scalability and extensibility in diverse deployment scenarios. These tests assess ACE-GNN’s adaptability to different levels of heterogeneity and system scale, as commonly encountered in edge environments. Specifically, we evaluate: (1) support for heterogeneous models, hardware, and tasks; (2) scalability with increasing numbers of edge devices; and (3) predictor generalization to unseen models, hardware platforms, and large-scale settings. Results show that ACE-GNN maintains robust performance across varied dynamic conditions.

\begin{figure}[t]
    \centering
    \includegraphics[width = 0.9\linewidth]{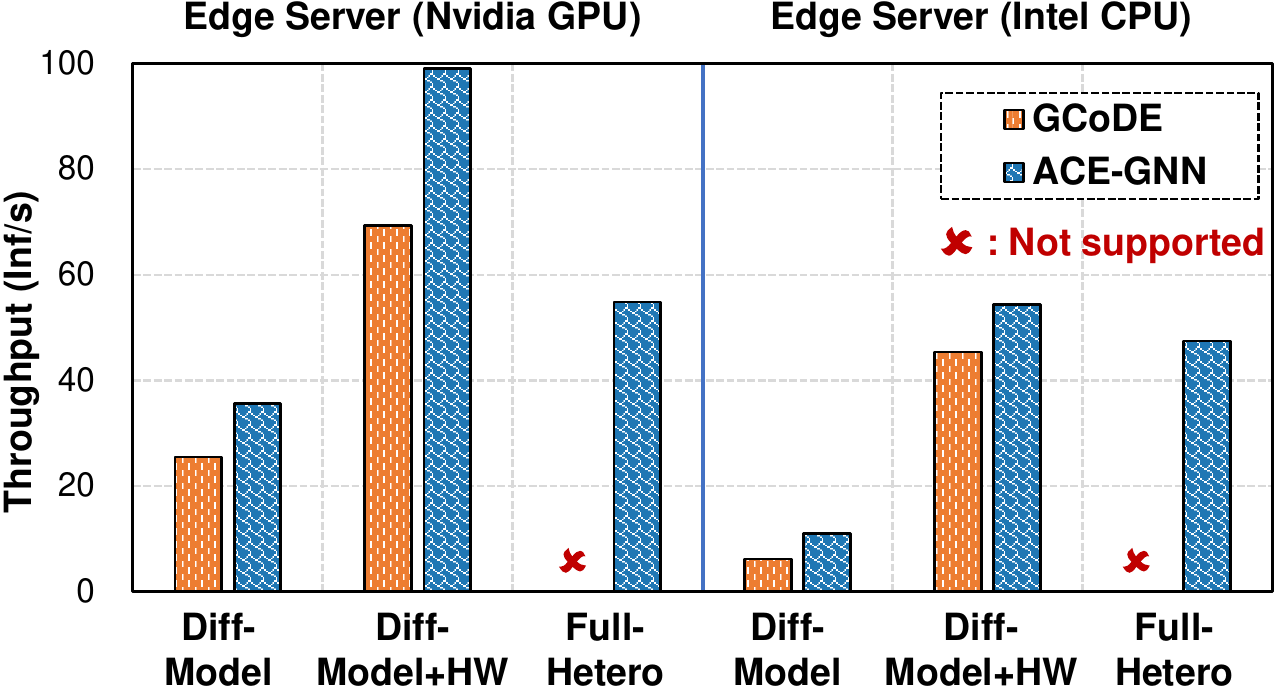}
    \caption{System throughput under various heterogeneous deployment scenarios.}
    \label{fig:hetero}
\end{figure}

\textbf{Extensibility to Heterogeneous Models, Tasks, and Devices.}
Fig.~\ref{fig:hetero} shows the performance of ACE-GNN under various heterogeneous deployment scenarios. In the \textit{Diff-Model} setting, all edge devices are Pi4B executing the same point cloud task but using different models—DGCNN and a model from GCoDE. ACE-GNN handles this model-level heterogeneity effectively, achieving up to $1.8\times$ higher throughput than GCoDE. In the \textit{Diff-HW+Model} setting, both hardware and models vary: Jetson TX2, Jetson Nano, Pi4B, and Pi3B each run GCoDE-designed models adapted to their hardware. Even with this heterogeneity, ACE-GNN achieves up to $1.4\times$ higher throughput. In the \textit{Full-Hetero} scenario, four different edge devices run different models and tasks—e.g., DGCNN on ModelNet40, GAT on Yelp, GCN on SIoT, and a GCoDE model on MR. ACE-GNN sustains strong performance with 50 Inf/s throughput, while GCoDE fails due to predictor and framework limitations. These results highlight ACE-GNN’s extensibility and adaptability across diverse heterogeneous settings.

\begin{figure}[t]
    \centering
    \includegraphics[width = 1\linewidth]{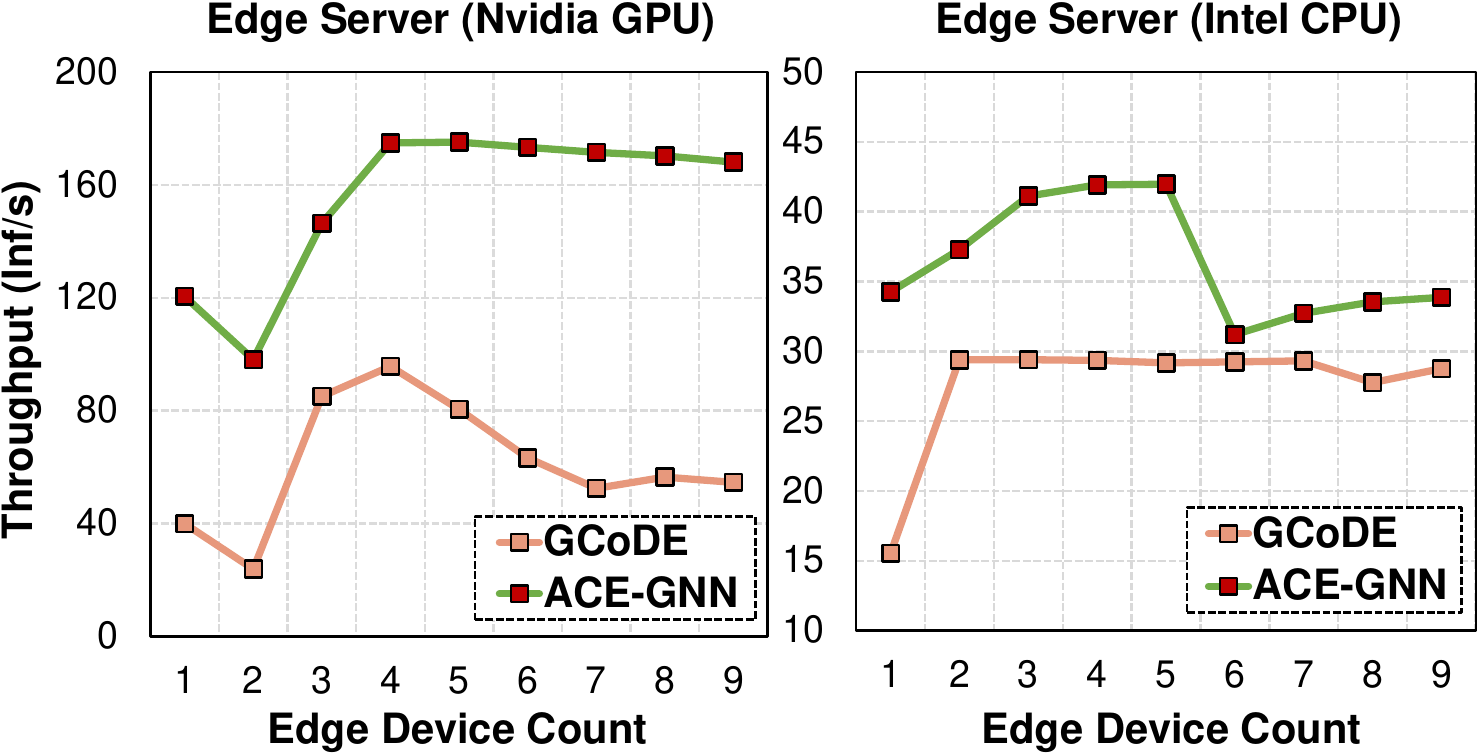}
    \caption{System throughput under larger-scale device deployments.}
    \label{fig:more_device}
\end{figure}

\textbf{Scalability with Device Count.}
We evaluate the scalability of ACE-GNN using up to 9 edge devices (5 Pi4B and 4 Pi3B) concurrently collaborating with the edge server.
As shown in Fig.~\ref{fig:more_device}, ACE-GNN maintains high throughput under increasing system scale.
The improvement is more pronounced when using an Nvidia GPU as the edge server, which handles high concurrency more efficiently than the Intel CPU.
In this setting, ACE-GNN achieves up to 3.1$\times$ higher throughput compared to GCoDE.

\textbf{Predictor Generalization to Unseen Settings.}
To evaluate predictor robustness, we generate new GNN architectures not seen during training or validation. Using these architectures, we construct over 100 test pairs and deploy them on real devices to obtain ground-truth latency. On these unseen models, the predictor achieves an average accuracy of $86.0\%$. It also maintains $89.3\%$ accuracy on a new hardware platform (Rockchip RK3588), and $96.4\%$ accuracy when scaling to 9 Raspberry Pi devices. These results demonstrate the predictor’s generalization capability across diverse, previously unseen deployment scenarios.

\begin{figure}[t]
    \centering
    \includegraphics[width = 1\linewidth]{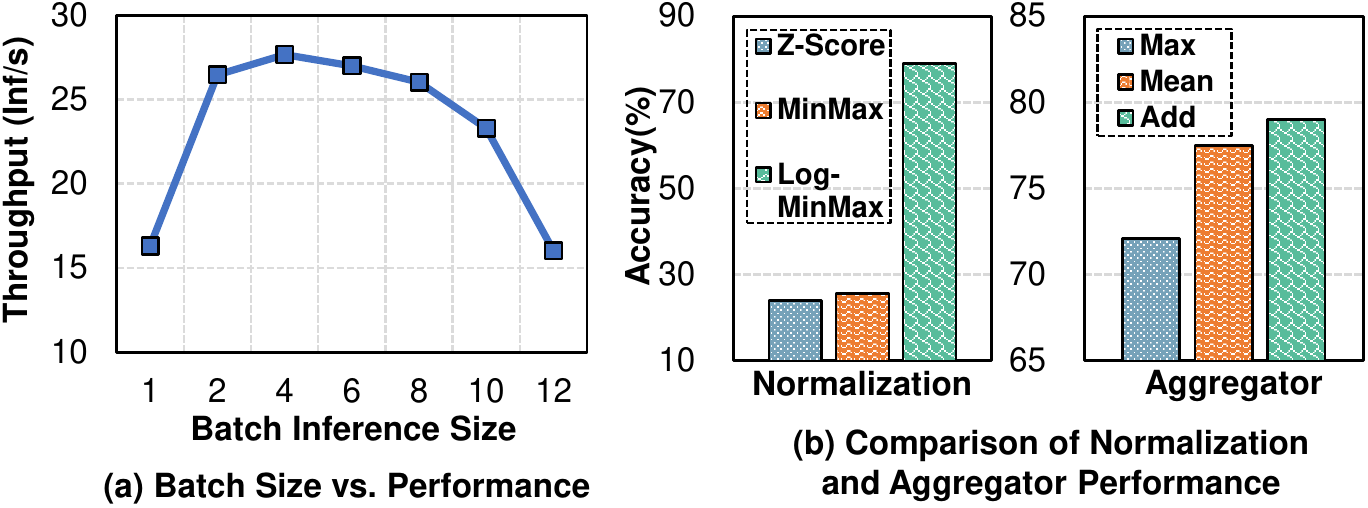}
    \caption{(a) Throughput comparison with varying batch sizes. (b) Accuracy comparison of normalization methods and aggregator settings.}
    \label{fig:ablation}
    \vspace{-6pt}
\end{figure}

\subsection{Ablation Study}\label{sec:ablation}

\textbf{Impact of Batch Inference Size.}
In practice, the batch inference strategy can enhance edge server resource utilization through appropriate batch size selection.
Fig.~\ref{fig:ablation} (a) shows the variation in batch inference performance with different batch sizes, using the DGCNN model and the ModelNet40 dataset.
As batch size increases, system throughput initially rises and then drops, emphasizing the benefits of batch processing and the need to limit its size to prevent severe edge workload.
In practice, a suitable batch limit can be determined by collecting GNN inference data with different batch sizes on the edge server during the data pre-collection.
Specifically, this information can be acquired in advance by running ACE-GNN's batch inference engine test on the target edge server.

\textbf{Impact of Normalization and Aggregator in Predictor.}
As shown in Fig.~\ref{fig:ablation} (b), due to the performance scale variation of samples in complex edge systems, both \textit{Z-Score} and \textit{MinMax} fail to properly balance the data distribution.
Conversely, the \textit{Log-MinMax} method we adopted effectively addresses this issue and significantly enhances predictive performance.
Moreover, the \textit{add} aggregator consolidates performance data from all devices, improving prediction accuracy.

\section{Conclusions}\label{sec:conclusions}

This paper proposes ACE-GNN, the first adaptive GNN co-inference framework tailored for dynamic edge environments. 
ACE-GNN addresses multiple GNN co-inference challenges in edge scenarios, including complex system awareness with multi-device access, novel co-inference techniques, specialized inference engine and communication middleware, and runtime optimization, achieving $12.7\times$ speedup and $11.7\times$ higher energy efficiency compared to SOTA frameworks GCoDE and Fograph.
By adaptively scheduling PP and DP, ACE-GNN improves GNN co-inference system stability and efficiency in dynamic edge environments.
Furthermore, ACE-GNN's system-level relative performance awareness method overcomes the limitations of existing approaches in accuracy and scalability, and reduces the training sample overhead.
We believe that ACE-GNN has the potential to promote broader application and exploration of GNN in edge scenarios.

\bibliographystyle{IEEEtran}
\small
\begingroup
\bibliography{ref.bib}
\endgroup

\vspace{-10mm}

\begin{IEEEbiography}[{\includegraphics[width=1in,height=1.25in,clip,keepaspectratio]{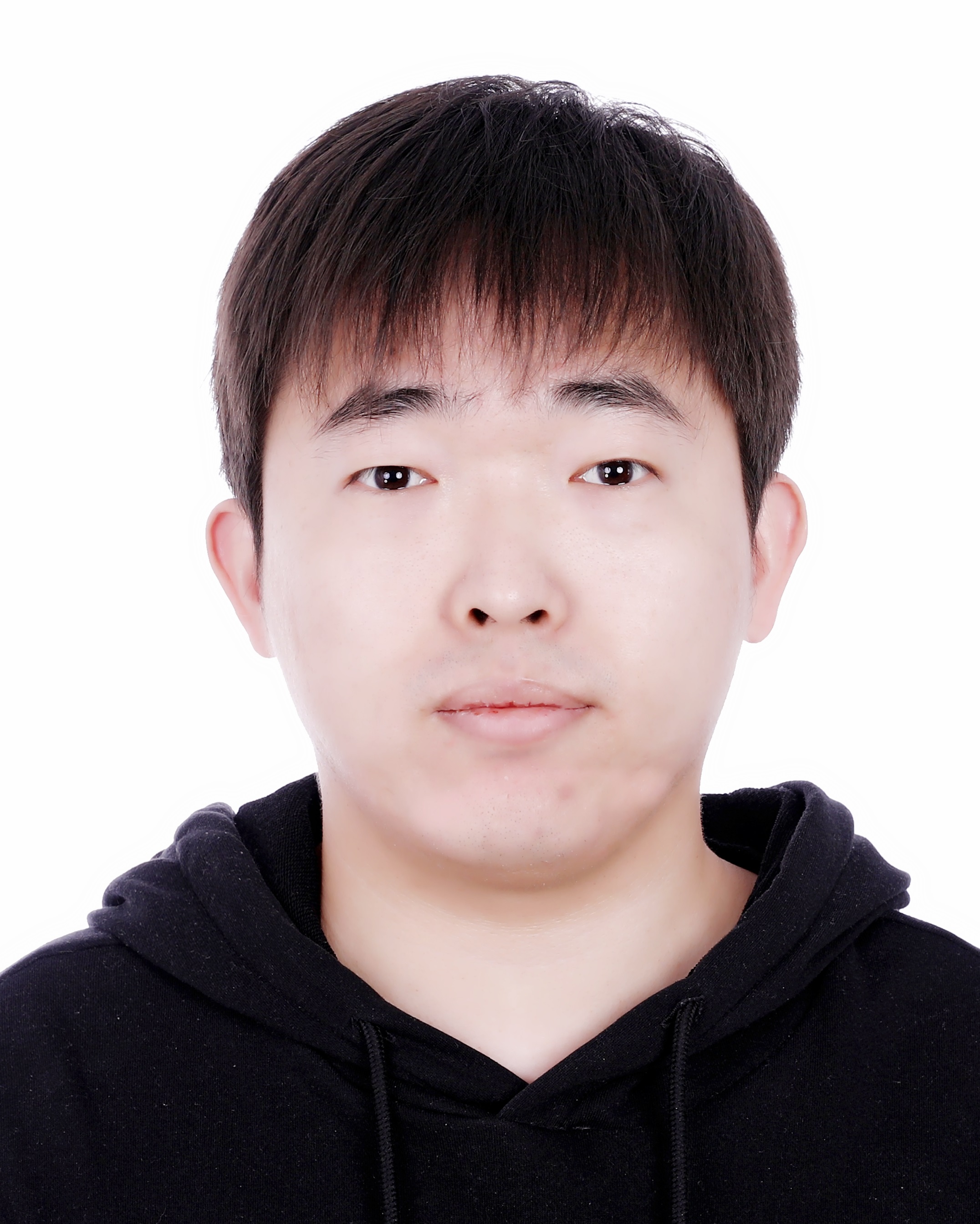}}]{Ao Zhou}

received the B.S. and M.S. degree in software engineering from Beijing University of Technology, Beijing, China, in 2018 and 2021, respectively, and the Ph.D. degree in software engineering from Beihang University, Beijing, China, in 2025. He is currently a Postdoctoral Researcher at the School of Software, Beihang University. His research interests include GNN acceleration, computer architecture, FPGA accelerator, and heterogeneous computing. He is one of the contributors to the popular GNN computation framework PyG.

\end{IEEEbiography}


\begin{IEEEbiography}[{\includegraphics[width=1in,height=1.25in,clip,keepaspectratio]{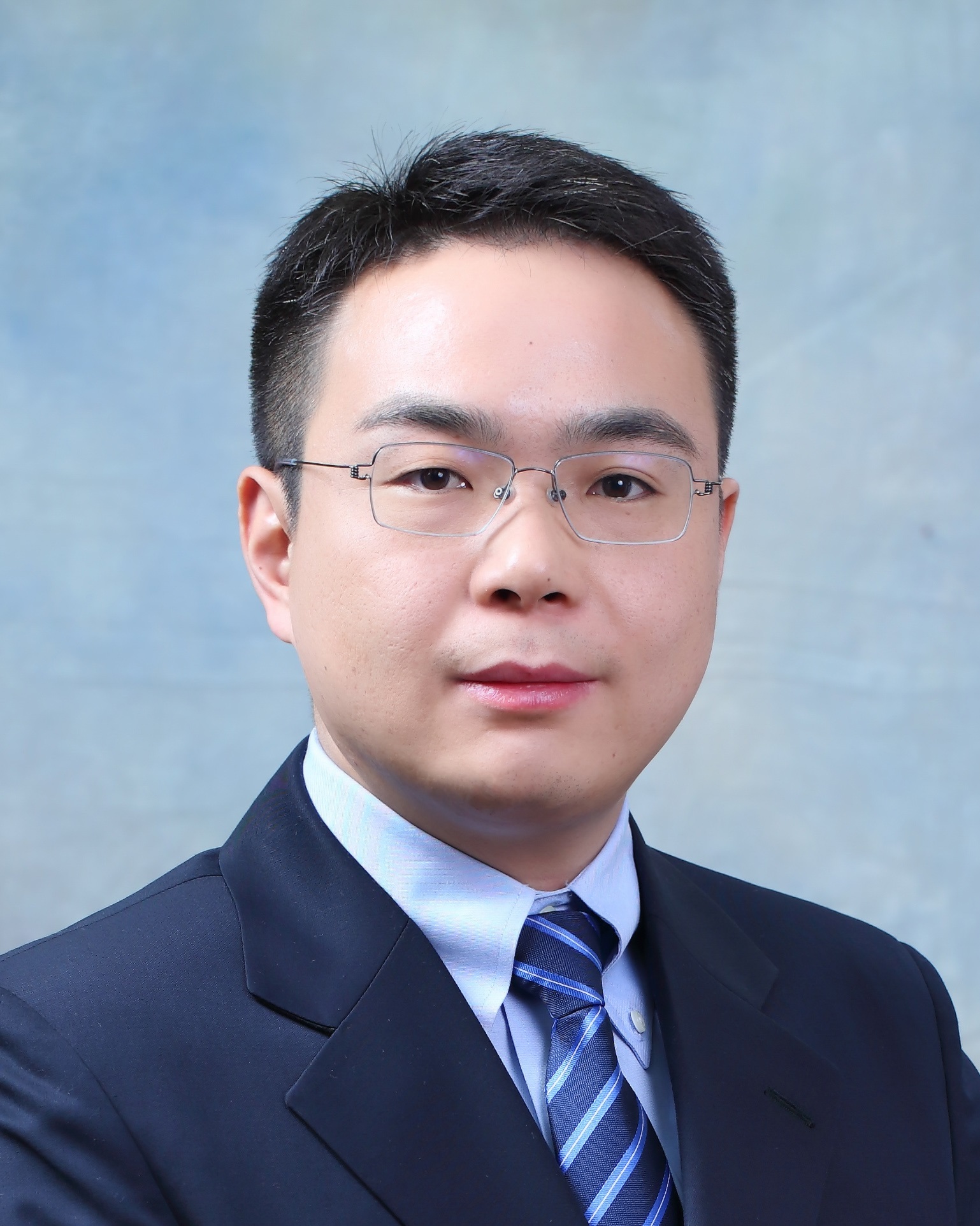}}]{Jianlei Yang}

(S'11-M'14-SM'20) received the B.S. degree in microelectronics from Xidian University, Xi'an, China, in 2009, and the Ph.D. degree in computer science and technology from Tsinghua University, Beijing, China, in 2014.

He is currently a Professor in Beihang University, Beijing, China, with the School of Computer Science and Engineering. From 2014 to 2016, he was a post-doctoral researcher with the Department of ECE, University of Pittsburgh, Pennsylvania, USA.
His current research interests include emerging computer architectures, hardware-software co-design and machine learning systems.

Dr. Yang was the recipient of the First/Second place on ACM TAU Power Grid Simulation Contest in 2011 and 2012. He was a recipient of IEEE ICCD Best Paper Award in 2013, ACM GLSVLSI Best Paper Nomination in 2015, IEEE ICESS Best Paper Award in 2017, ACM SIGKDD Best Student Paper Award in 2020.

\end{IEEEbiography}

\vspace{-15mm}

\begin{IEEEbiography}[{\includegraphics[width=1in,height=1.25in,clip,keepaspectratio]{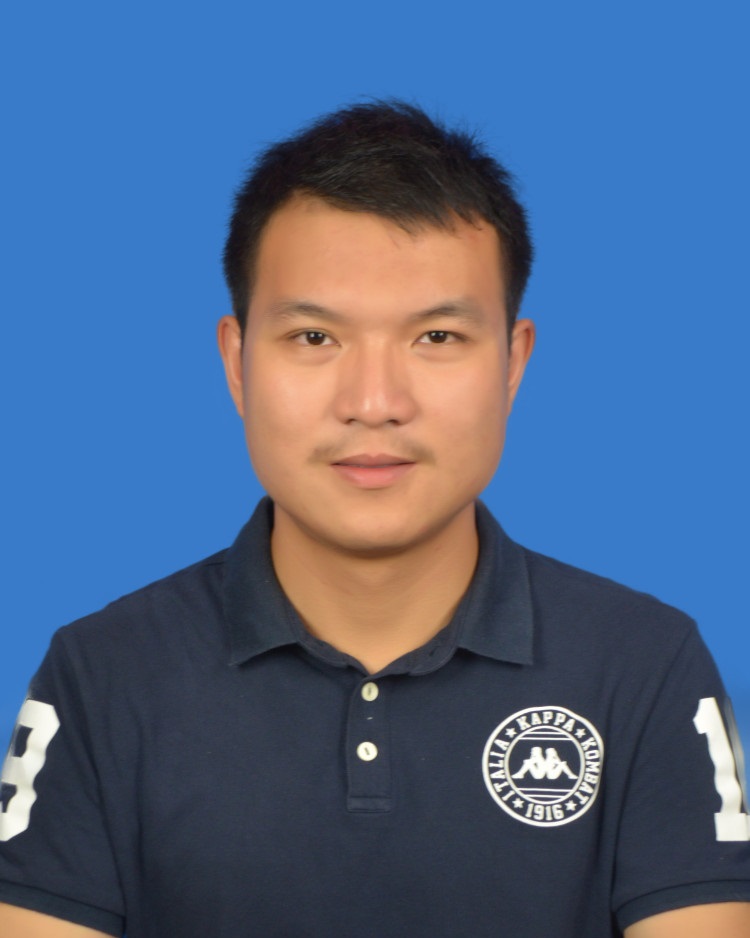}}]{Tong Qiao}

received the B.S. degree in computer science and technology from Beihang University, Beijing, China, in 2020. He is currently pursuing the Ph.D. degree at the School of Computer Science and Engineering, Beihang University, China. His research interests include graph neural networks acceleration, and system for machine learning.

\end{IEEEbiography}

\vspace{-12mm}

\begin{IEEEbiography}[{\includegraphics[width=1in,height=1.25in,clip,keepaspectratio]{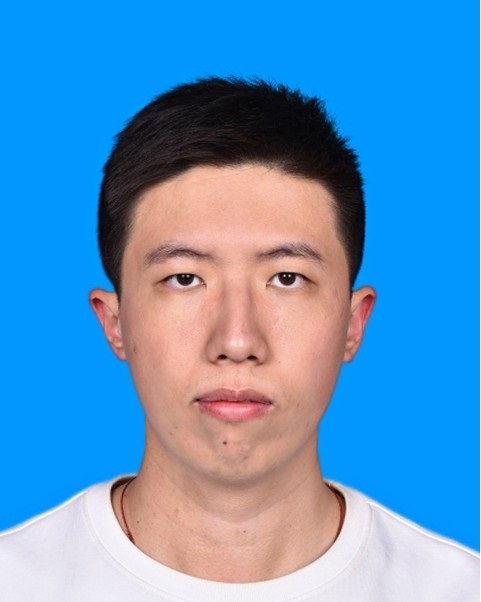}}]{Yingjie Qi}

received the B.S. degree in computer science and technology from Beihang University, Beijing, China, in 2020. He is currently pursuing the Ph.D. degree at the School of Computer Science and Engineering, Beihang University, China. His research interests include graph neural networks acceleration, processing-in-memory architectures and deep learning compilers.

\end{IEEEbiography}

\vspace{-15mm}

\begin{IEEEbiography}[{\includegraphics[width=1in,height=1.25in,clip,keepaspectratio]{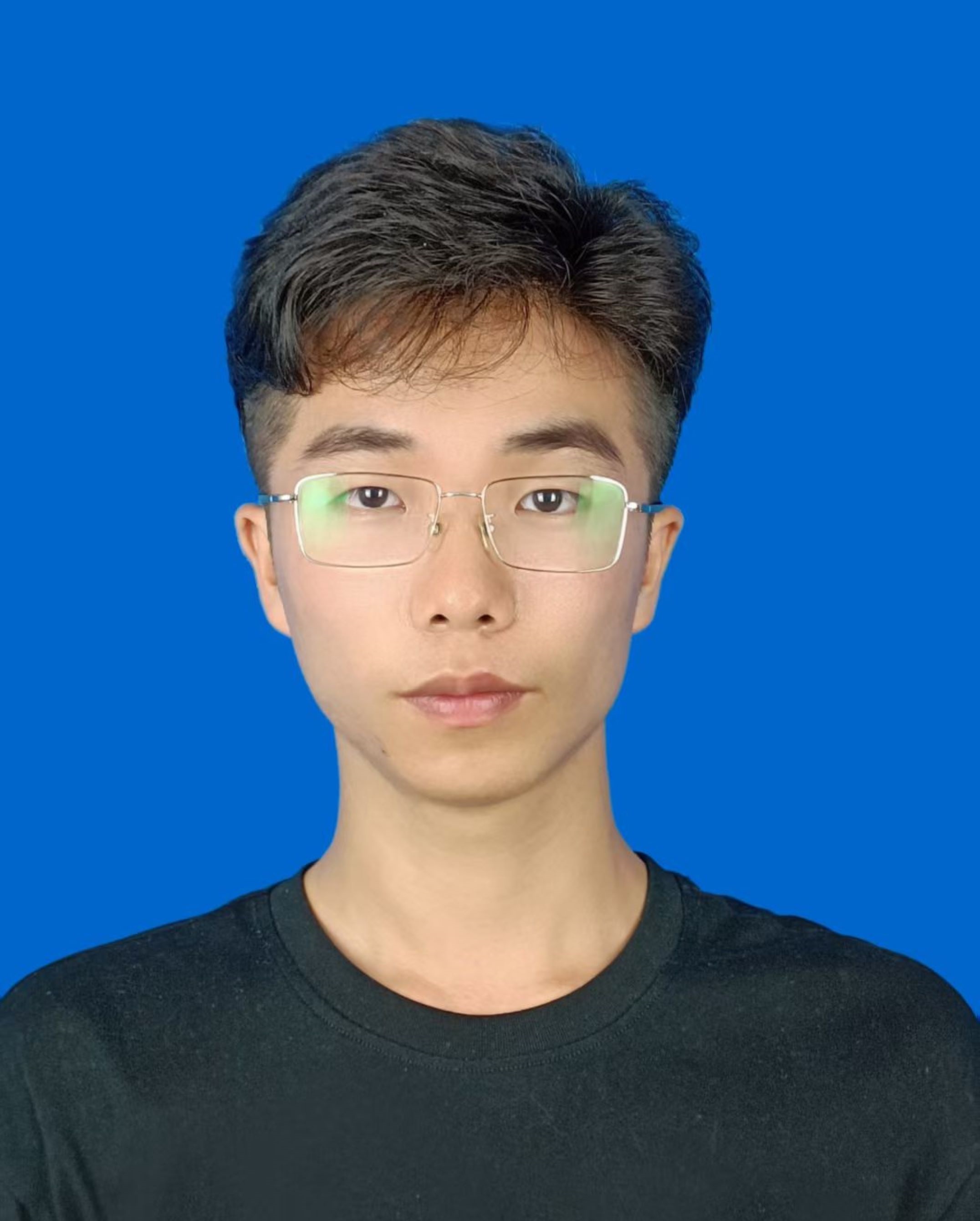}}]{Xinming Wei}

is currently pursuing his Bachelor's degree in Computer Science and Technology at Beihang University, Beijing, China, expected to graduate in 2026.
His recent research interests focus on visual computing and edge computing.
As an undergraduate researcher, he is actively exploring innovative approaches to optimize computational efficiency  in distributed systems.

\end{IEEEbiography}

\vspace{-15mm}

\begin{IEEEbiography}[{\includegraphics[width=1in,height=1.25in,clip,keepaspectratio]{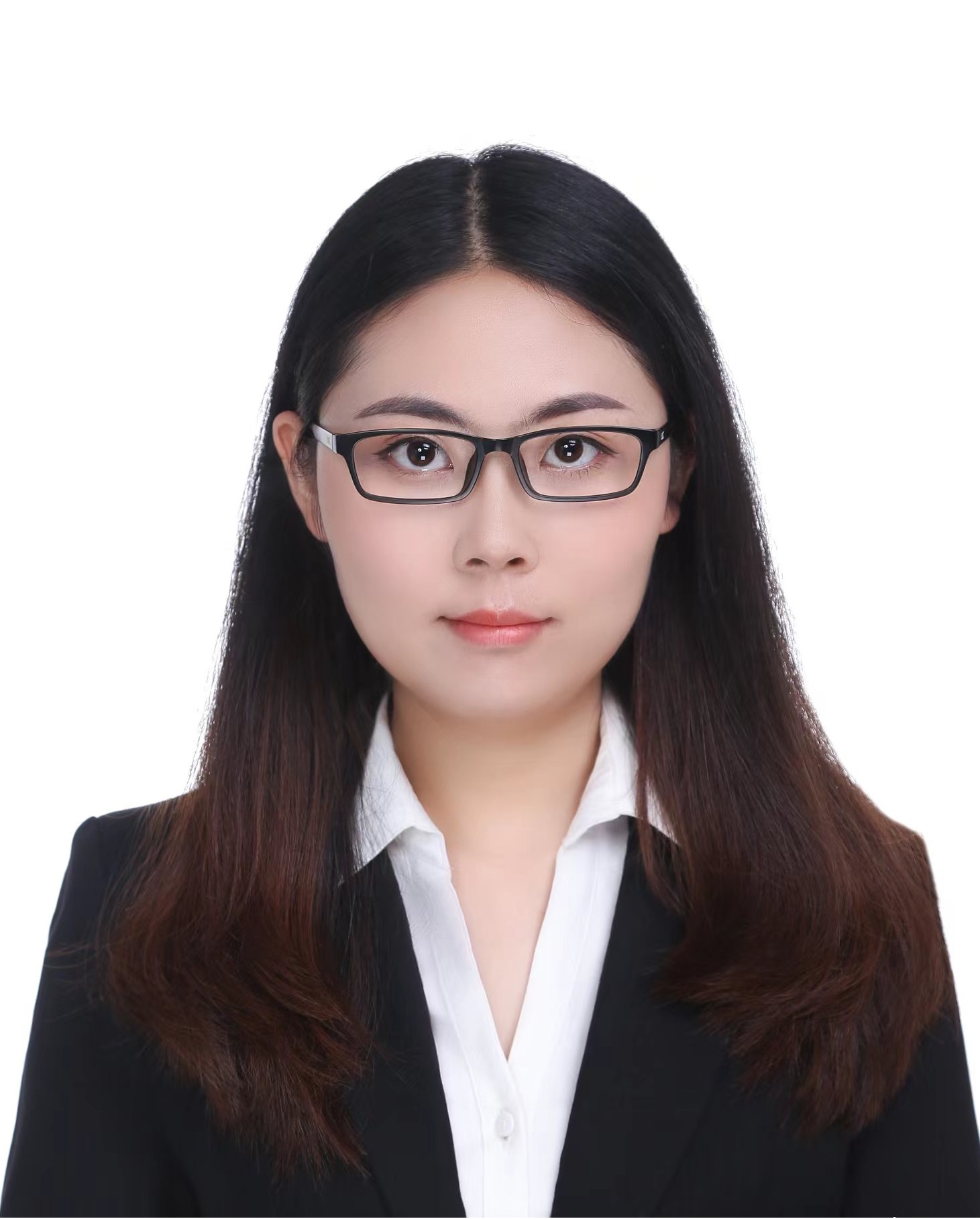}}]{Cenlin Duan}

received the B.S. degree in electronic science and technology from University of Electronic Science and Technology of China, Chengdu, China, in 2015, and the M.S. degree in software engineering from Xidian University, Xi'an, China, in 2018. She is currently pursuing the Ph.D. degree at the School of Integrated Circuit Science and Engineering, Beihang University, Beijing, China. Her current research interests include processing-in-memory architectures and deep learning accelerators.

\end{IEEEbiography}

\vspace{-15mm}
\begin{IEEEbiography}[{\includegraphics[width=1in,height=1.25in,clip,keepaspectratio]{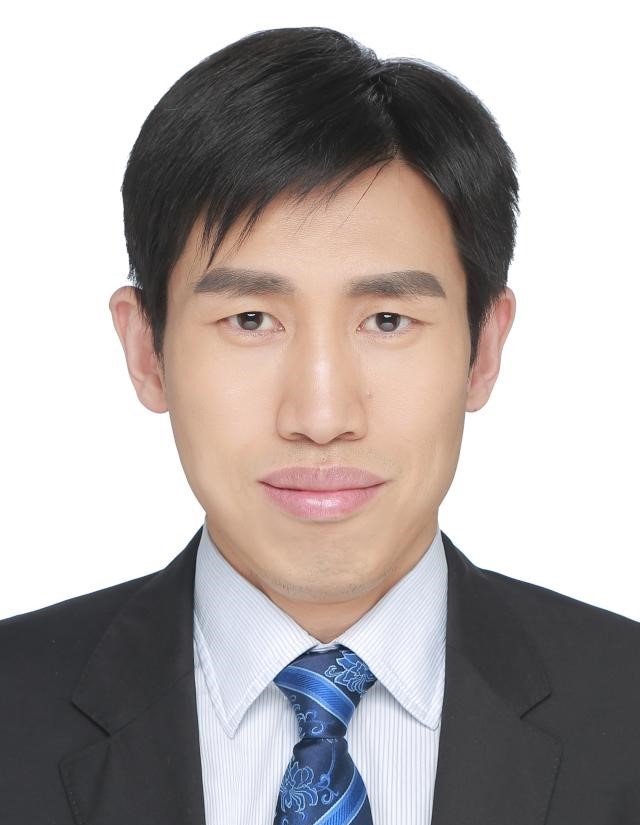}}]{Weisheng Zhao}

(Fellow, IEEE) received the Ph.D. degree in physics from the University of Paris Sud, Paris, France, in 2007.

He is currently a Professor with the School of Integrated Circuit Science and Engineering, Beihang University, Beijing, China. In 2009, he joined the French National Research Center, Paris, as a Tenured Research Scientist. Since 2014, he has been a Distinguished Professor with Beihang University. He has published more than 200 scientific articles in leading journals and conferences, such as \textit{Nature
Electronics}, \textit{Nature Communications}, \textit{Advanced Materials}, IEEE Transactions, ISCA, and DAC. His current research interests include the hybrid integration of nanodevices with CMOS circuit and new nonvolatile memory (40-nm technology node and below) like MRAM circuit and architecture design.

Prof. Zhao was the Editor-in-Chief for the {\sc{IEEE Transactions on Circuits and System I: Regular Paper}} from 2020 to 2023.

\end{IEEEbiography}

\vspace{-15mm}
\begin{IEEEbiography}[{\includegraphics[width=1in,height=1.25in,clip,keepaspectratio]{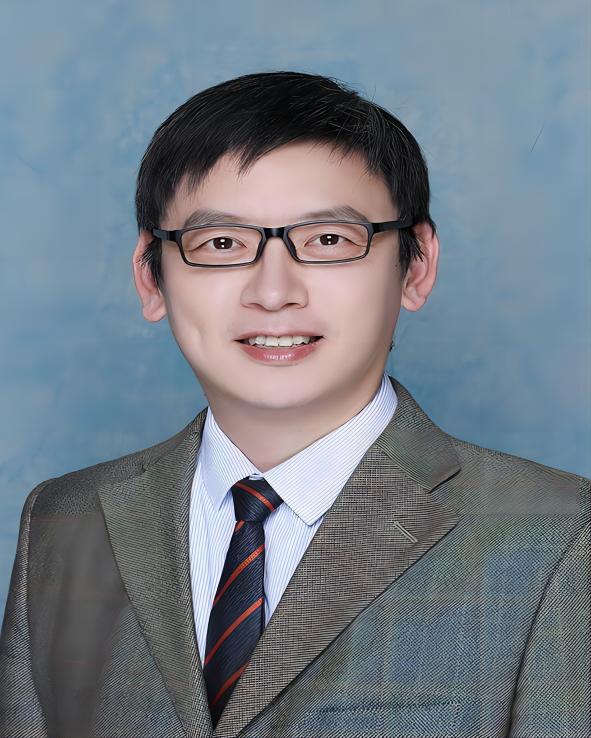}}]{Chunming Hu}

received the PhD degree in computer science and technology from Beihang University, Beijing, China, in 2006.

He is currently a professor and dean of the School of Software, Beihang University, Beijing, China. His current research interests include distributed systems, system virtualization, mobile computing and cloud computing.

Prof. Hu is currently one of the W3C Board of Directors, and serving as the Deputy Director of W3C China.

\end{IEEEbiography}

\end{document}